\documentclass[a4paper]{PoS}
\title{Quantum Finite Elements for  Lattice Field Theory}

\ShortTitle{Quantum Finite Elements}

\author{
Richard C. Brower$^*$\\
Boston University, Boston, MA 02215, USA\\
\email{email: brower@bu.edu}}
				
\author{George Fleming$^*$\\
			  Yale University, Sloane Laboratory, New Haven, CT 60520,\\
				\email{email: george.fleming@yale.edu}}

\author{Andrew Gasbarro$^*$\\
       Yale University, Sloane Laboratory, New Haven, CT 60520,\\
       \email{email: andrew.gasbarro@yale.edu}}
       
\author{Timothy Raben\\
      	Brown University, Providence, RI, 02912\\
       \email{email: timothy\_raben@brown.edu}}
       
\author{Chung-I Tan\\
      	Brown University, Providence, RI, 02912\\
       \email{email: chung-i\_tan@brown.edu}}
       
\author{Evan Weinberg\\
        Boston University, Boston, MA 02215, USA\\
       \email{email: weinbe2@bu.edu}}

\RequirePackage{graphicx}
\usepackage{amsmath}
\usepackage{amssymb}
\usepackage{cite}
\usepackage{bm}
\usepackage{caption}
\usepackage{slashed}
\usepackage{mathtools}
\usepackage{wrapfig}
\usepackage{braket}

\newcommand{\ignore}[1]{}
\newcommand{\be}{\begin{equation}}
  \newcommand{\bea}{\begin{eqnarray}}
  \newcommand{\eea}{\end{eqnarray}}
  \newcommand{\beq}{\begin{equation}}
  \newcommand{\ee}{\end{equation}}
  \newcommand{\eeq}{\end{equation}}
  
  \newcommand{\<}{\langle\,}

  \newcommand{\half}{{\frac{1}{2}}}
  \renewcommand{\>}{\rangle}
\newcommand\dd{\partial}
\newcommand\nn{\nonumber \\}

\def\32{{\frac{3}{2} } }

\FullConference{The 33rd International Symposium on Lattice Field Theory\\
		14 -18 July 2015\\
		Kobe International Conference Center, Kobe, Japan*}

              \abstract{Viable non-perturbative methods for lattice
                quantum field theories on curved manifolds are
                difficult. By adapting features from the traditional
                finite element methods (FEM) and Regge Calculus, a new
                simplicial lattice Quantum Finite Element (QFE)
                Lagrangian is constructed for fields on a smooth
                Riemann manifold.  To reach the continuum limit
                additional counter terms must be constructed to cancel
                the ultraviolet distortions.  This is tested by the
                comparison of phi 4-th theory at the Wilson-Fisher
                fixed point with the exact Ising (c =1/2) CFT on a 2D
                Riemann sphere. The Dirac equation is also constructed
                on a simplicial lattice approximation to a Riemann
                manifold by introducing a lattice vierbein and spin
                connection on each link. Convergence of the QFE Dirac
                equation is tested against the exact solution for the
                2D Riemann sphere. Future directions and applications
                to Conformal Field Theories are suggested.
                \\
                \rule{4cm}{.2pt}\\
                $^*$ Speakers}

\begin{document}

%%%%%%%%%%%%%%%%%%%%%%%%%%%%%%%%%%%%%%%%%%%%%%%%%%%%%%%%%%%%%%%%%%%%%%%%%%%%%%%%%%%%%%%%%%%%%%%%%%%%%%%%%%
\section{Introduction}

Lattice gauge theory on hypercubic Euclidean lattices provides a
powerful {\em ab initio} approach to strongly coupled field
theories with increasingly accurate predictions for lattice Quantum
Chromodynamics. However there are applications in quantum field theory
and condensed matter physics that would benefit from
equally powerful methods on curved manifolds.  One example is the
recent 
suggestion to apply lattice methods to the radial quantization of conformal field
theories~\cite{Brower:2012vg, Brower:2014gsa, Brower:2015zea}. 

In  radial quantization, the  conformal field theory
in flat Euclidean space  $\mathbb{R}^D$ is mapped to $\mathbb{R}\times
\mathbb{S}^{D-1}$,
\be
ds^2_{flat} =dx^{\mu}dx^{\mu} = r^2_0 e^{2t}(dt^2 + d\Omega^2_{D-1})
\xrightarrow{Weyl} (dt^2 + d\Omega^2_{D-1}) \; .
\label{eq:Radial}
\ee
In the absence of a conformal anomaly the Weyl rescaling,
which drops the factor, $r^2/r^2_0 = e^{\textstyle 2 t}$, is allowed.
The new geometry is a cylinder with transverse 
spheres, $\mathbb{S}^{D-1}$, with unit radius.  The full Euclidean conformal group $SO(D+1,1)$ is in one to one
correspondence with the isometries of $AdS^{D+1}$  and in global
co-ordinates  the radial
dual CFT lives on the $\mathbb{R}\times
\mathbb{S}^{D-1}$  boundary (\ref{eq:Radial}) of anti-de Sitter space.    Instead of the more
conventional Hamiltonian for Euclidean time  evolution, in radial
quantization translations on   new ``time''  axis, $t = \log(r/r_0)$,  along the cylinder are generated by the 
Dilation operator whose discrete CFT spectra are the dimension of
the operators in the  operator product representation. This is just
one of many application of non-perturbative quantum field theory on curved
manifolds. 

Relative to a flat Euclidean manifold, ${\mathbb R}^D$, the technical problem
on curved manifolds is the lack of an infinite sequence of regular
crystallographic lattices approaching the continuum, thus compounding the problem of
renormalization and symmetry restorations as the cut-off is removed.  For example on the two sphere, the 
icosahedral group is the largest discrete subgroup of rotations. At
present our numerical tests are restricted to the 2D
Riemann sphere $\mathbb S^2$ but these methods should applied more generally to  lattice quantum field theory on
a smooth crurved Riemann manifolds.

We introduce a sequence of simplicial lattices to conform to 
the target Riemann manifold. The
literature provides some tools for addressing these problems.
In  Regge Calculus ~\cite{Regge:1961px} a lattice Einstein action of
General Relativity seeks to define a non-perturbative quantum
theory as a sum over simplicial geometries.  A similar
formalism was also developed in Euclidean  flat space
in a remarkable series of papers by  Christ, Friedberg and Lee~\cite{Christ:1982zq,Christ:1982ck,Christ:1982ci} on
random lattice gauge field
theory.  Both of these focus on the use of
an ensemble of   lattices that are conjectured to restore symmetries by averaging
over this ensemble. In a seemingly separate history, finite element
methods (FEM) on simplicial lattices have been
developed in order to  solve differential equations on
smooth manifolds.  The FEM approach provides a strong theoretical
framework~\cite{StrangFix200805} in which convergence is
guaranteed for  solutions to the classical equations of
motion, for a properly constrained sequence of simplicial lattice
refinements. In this  proceedings, we report on a new approach, borrowing the geometry of Regge
Calculus or equivalently the discrete exterior calculus  while seeking to restore the continuum limit by a
fixed sequence of simplicial lattices as in the application of  FEM to
partial differential equations. This extension  to quantum filed
theory on curved Riemann manifolds, we will be refer to as {\bf Quantum Finite Elements (QFE)}.  

Here we focus on two issues for QFE: First,  the necessity of introducing simplicial counter terms
to remove the ultraviolet distortions, as the cut-off is removed. We
require that  the  simplicial lattice quantum field theory  converges exactly
to the continuum quantum field theory~\cite{Brower:2015zea}. Second, the proper  treatment of a Dirac field on a
simplicial manifolds~\cite{BrowerDirac}.  Both of these topics will be
presented in  more detail in  future publications.  To explore the first topic, we develop the QFE formalism for the scalar 
$\phi^4$ theory, which is conformal at the Wilson-Fisher fixed point.
We expose the inadequacy of the conventional  FEM Lagrangian due to
ultraviolet divergences and find a quantum couter term for the
QFE Lagrangian that solves this problem.   For the second we
construct  lattice Fermions on a simplicial lattice providing
an explicit construction of lattice vierbein and spin connection
essential to properly account for the curvature of the target Riemann
manifold. Numerical tests of convergence to the continuum
limit for problmes are provided for the case of the 2D Riemann sphere $\mathbb S^2$. 
We suggest future directions to generalize these methods
to interacting lattice theories with scalars, Fermions and non-Abelian
gauge fields. The challenges are profound but we are encouraged by progress to date.

\section{Scalar Fields on Simplicial Lattice}

The scalar $\phi^4$ theory provides a natural first step to developing a QFE
lattice action. The construction proceeds in two steps:
%\begin{enumerate}[noitemsep, topsep=0pt]
\begin{enumerate}
\item First replace the smooth Riemann manifold (${\cal    M}, g$) by an approximating
  piecewise flat manifold (${\cal  M}_\sigma,g_\sigma$) composed of elementary simplices.  
\item  Second expand the field, $\phi(x)$,  in a finite element basis on each
  simplex: $\phi(x) \simeq W^i(x) \phi_i$.
\end{enumerate}
It is important to clearly separate these two steps. The first step is a
simplicial approximation to the geometry of manifold emphasized in the
Regge Calculus treatment~\cite{Regge:1961px} and the second step is
an approximation of the function space for the field into a local basis
emphasized in classical FEM theory~\cite{StrangFix200805}. Of course
both are necessary and have mutual interdependences required for
consistency and convergence to the continuum theory. Also we should
warn the reader, we find that both of these steps require significant
departures from earlier attempts.

The action for the scalar Lagrangian on a D-dimensional Euclidean
Riemann manifold is
\be
S = \half  \int_{\cal M}  d^Dx \sqrt{g}\; [g^{\mu\nu}\dd_\mu
\phi(x)\dd_\nu \phi(x) +  m^2 \phi^2(x) + \lambda \phi^4(x) ] \; ,
\label{eq:scalar}
\ee
where the proper distance on the manifold is given by the metric,
\be
ds^2 =g_{\mu \nu}(x) dx^\mu dx^\nu  \; .
\ee
Since the metric is a positive definite  symmetric tensor, it can be
Cholesky factored ( $G  =  E E^T$) by introducing a local tangent plane at each point $x^\mu$ with
orthonormal  co-ordinates,  $\vec y = y^1 \hat n_1 + y^2 \hat n_2
\cdots +  y^D \hat n_D $  and by expanding in differential~\footnote{In flat Euclidean space upper
and lower tangent plane indices are equivalent and they 
will often be  represented in vector notation with replacements
such as $e^a_\mu
  \rightarrow \vec e_\mu$. } (1-forms) 
$d y^a =  \frac{\partial y^a}{\partial x^\mu} dx^\mu  \equiv   e^a_\mu
dx^\mu$ so that
\be
 ds^2 = d\vec y \cdot d\vec y=  \vec e_\mu  \cdot \vec e_\nu \; dx^\mu
 dx^\nu  \; .
\ee

\paragraph{Map of Riemann Manifold ${\cal M}$  to Simplicial
  Complex ${\cal M}_\sigma$:}

\begin{figure}
\centering
  \includegraphics[width=.8\textwidth]{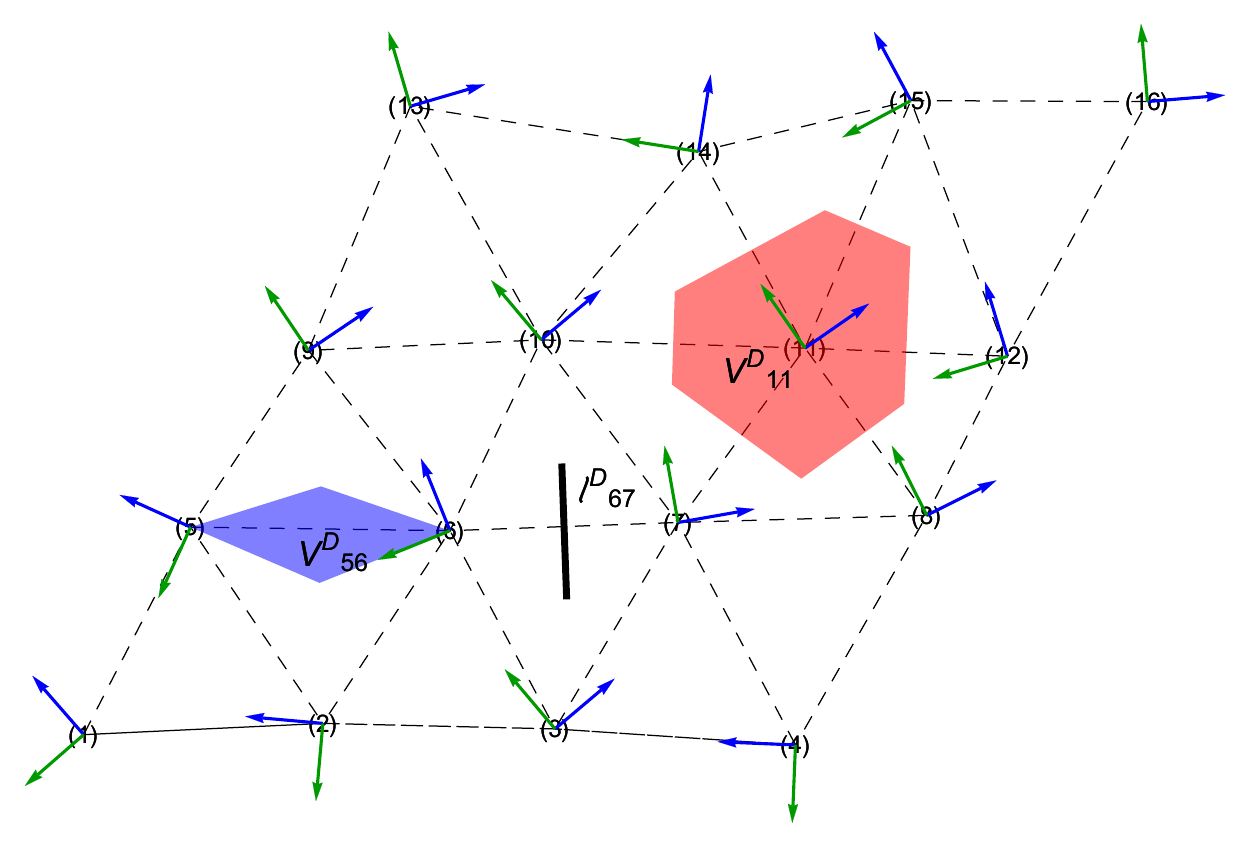}
\caption{\label{fig:lattice} A 2D simplicial complex with  points
  $(\sigma_0)$, edges ($\sigma_1$) and triangles $(\sigma_2)$. At each
  vertex, $\sigma_0$, is   a dual
  polytope in  $*\sigma_0$ ( illustrated in red) and at each link $\sigma_1$
a dual area  in $\sigma_1 \wedge *\sigma_1$ (illustrated in blue). The
arrows at each site represent a random basis for the
local tangent plane. }
\end{figure}

We need to construct a sequence of  manifolds (${\cal M}_\sigma,g_\sigma$)
composed of simplices
(triangles, tetrahedrons, ...  for D = 2, 3,...)  and assign distances
to the edges $l^2_{ij}$ and a flat metric in the interior of each
simplex.  The crucial step is to make a one to one correspondence
between  points on the smooth Riemann manifold (${\cal M},g$) and
points on the new piece-wise flat
simplicial manifold (${\cal M}_\sigma,g_\sigma$) that preserves distance to order $O(a^2)$ in the
diameter $a$ of the simplices.  This is not easy in general but 
one approach is to construct  a smooth isometric embedding of the D-dimensional Riemann
manifold (${\cal M},g$) into a higher dimensional Euclidean space
$\vec r \in {\mathbb R^N}$. For the $\mathbb S^D$ sphere this is
easily done with a $\vec r \in {\mathbb R^{D+1}}$ and the constraint $\vec
r \cdot \vec r =1$. Then one uses a Voronoi construction of simplices on
a set of discrete sites at $x = r_i$  assigning the embedded distances,
$l_{ij} = |r_i - r_j|$, to the edges, as illustrated in Fig.~\ref{fig:lattice} for $D
= 2$. Smoothness  should guarantee convergence of the simplicial
manifold (${\cal M}_\sigma,g_\sigma$) to the target manifold
(${\cal M},g$) as $a \rightarrow 0$. The best approach to control
$O(a^2)$ errors is an important consideration of course.  Other methods
based entirely on intrinsic  geometry would be useful.

The interior of each simplex (e.g. triangle, tetrahedron,...)  can then
be parameterized by $D+1$ barycentric co-ordinates, $0 \le\xi^i \le 1$,
\be
\vec y  = \xi^0 \vec r_0 + \xi^1 \vec r_1 + \cdots + \xi^{D} \vec
r_{D} = \sum^D_{i=1} \xi^i \vec l_i + \vec r_0 \; ,
\ee
with the constraint $\xi^1 + \xi^2  + \cdots + \xi^{D+1} =1$.
\begin{figure}[h]
\centering
  \includegraphics[width=.7\textwidth]{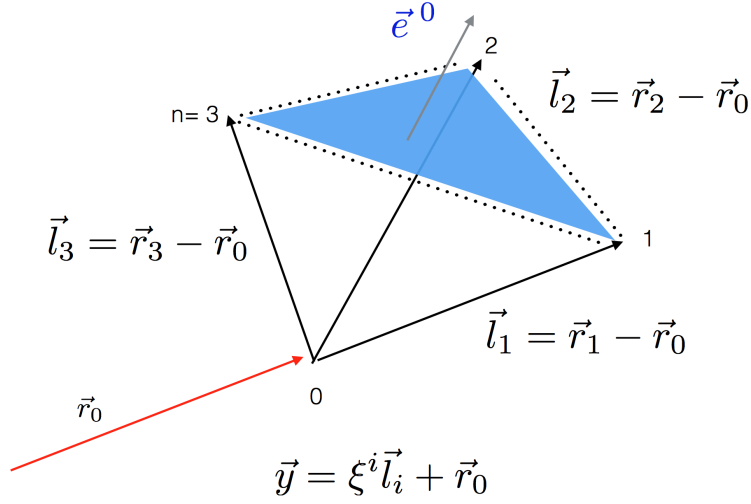}
\vskip -0.7 cm
\caption{\label{fig:simplex}  The n-simplex, illustrated for $n = 3$,
  can be defined by n edge vectors 
$ \vec l_i \equiv \vec l_{i0}  = \vec r_i - \vec r_0$, picking arbitrarily
the 0-th vertex. The remaining $n(n-1)/2$ edges are $\vec
l_{ij} = \vec l_i - \vec l_j$.}
\end{figure}
To pick   a unique co-ordinate system on ${\cal M}_\sigma$, we may
eliminate  one $\xi^i$. for example $\xi^0$ and set $\vec l_i = \vec
l_{i0} = \vec r_i - \vec
r_0$ so the flat $D\times D$  metric tensor is given by  $g_{ij} =  \dd_i\vec y \cdot \dd_j \vec y =
\vec l_i \cdot \vec l_j $.

 Now the action\footnote{Note on a flat simplex there is a single tangent plane
with $y^a$ providing  co-ordinates  for all points, $\xi^i$, in the
interior.}
on each simplex in  (${\cal  M}_\sigma,g_\sigma$)  is uniquely
determined by Eq.~\ref{eq:scalar} on this simplicial  Riemann manifold 
\bea
I_{\sigma} &=& \frac{1}{2} \int_{\sigma} d^D y [\vec \nabla \phi(y)  \cdot \vec
\nabla \phi(y) + m^2 \phi^2(y)  + \lambda \phi^4(y)] \nn
& =&  \frac{1}{2} \int_\sigma d^D\xi  \sqrt{g}\; [ g^{ij} 
\dd_i \phi(\xi)  \dd_j \phi^2(\xi) + m^2 \phi^2(\xi)  + \lambda
\phi^4(\xi)] \; ,
\label{eq:SimplicialAction}
\eea
where now $\sqrt{g} = \sqrt{\det(g_{ij})} = D!
V_D$ with $V_D$  volume of the D-simplex and $g^{ij}$ is the inverse of the metric tensor.

\begin{figure}[ht]
\centering
\includegraphics[width=0.32\textwidth]{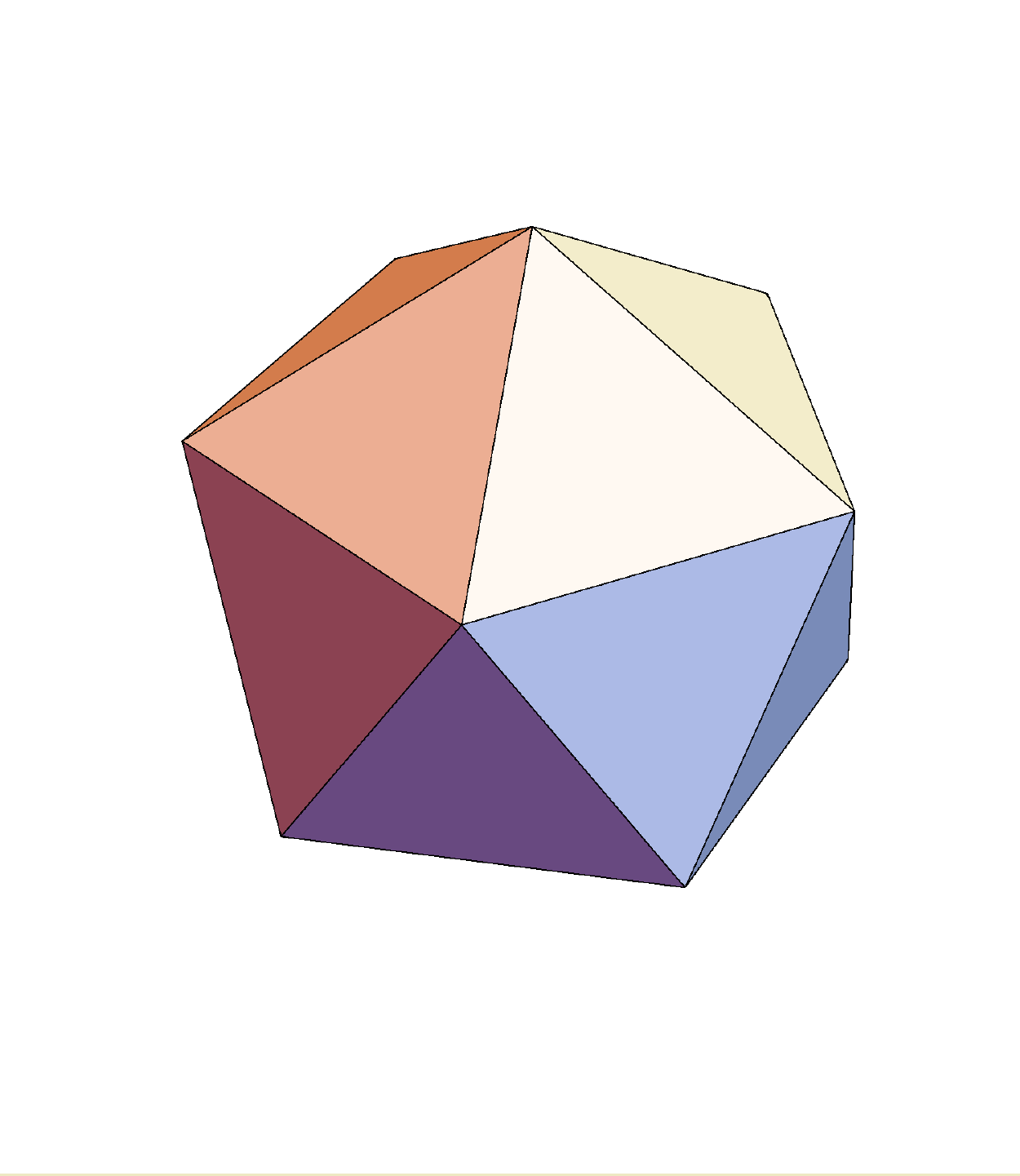}
\includegraphics[width=0.32\textwidth]{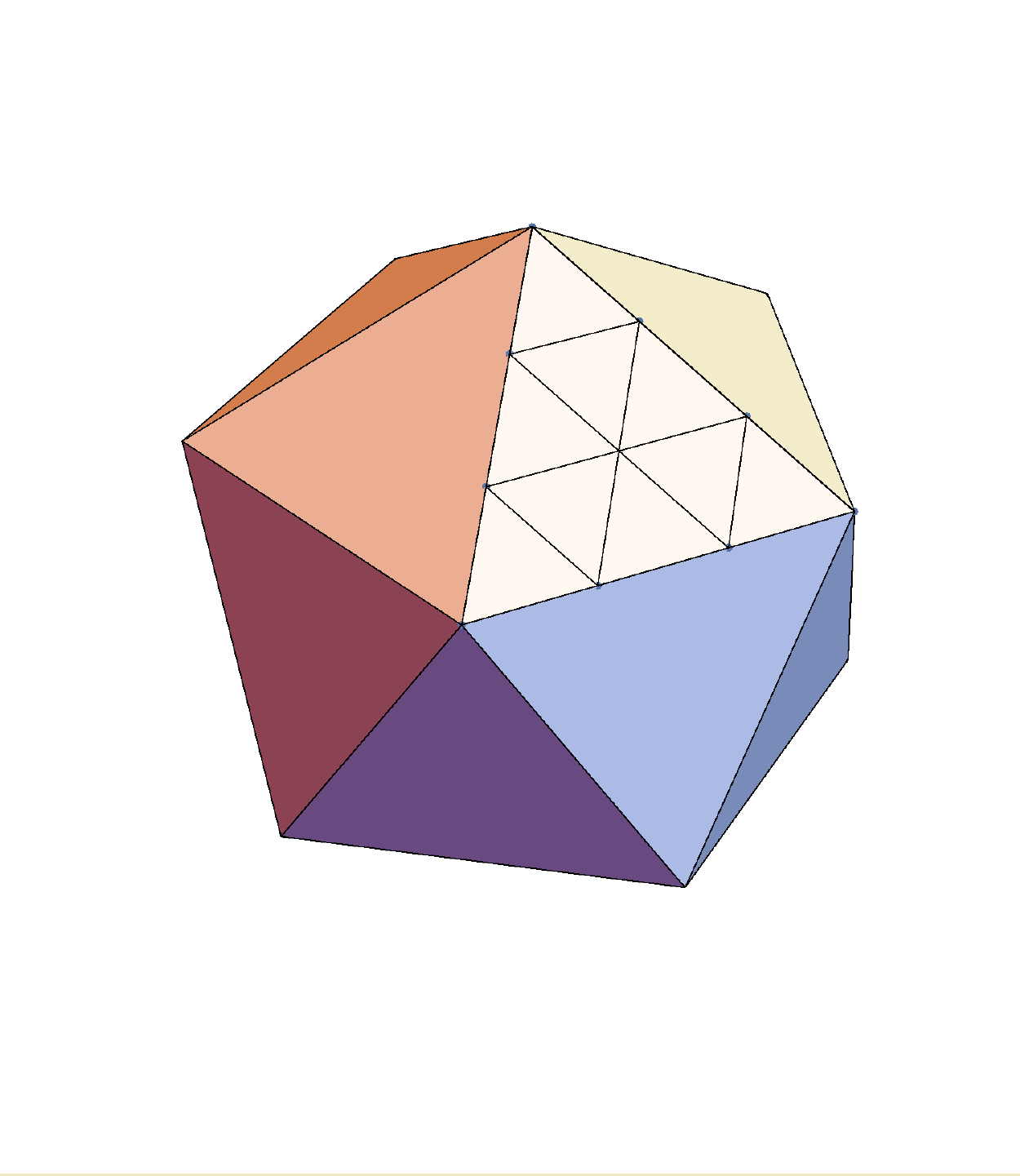}
\includegraphics[width=0.32\textwidth]{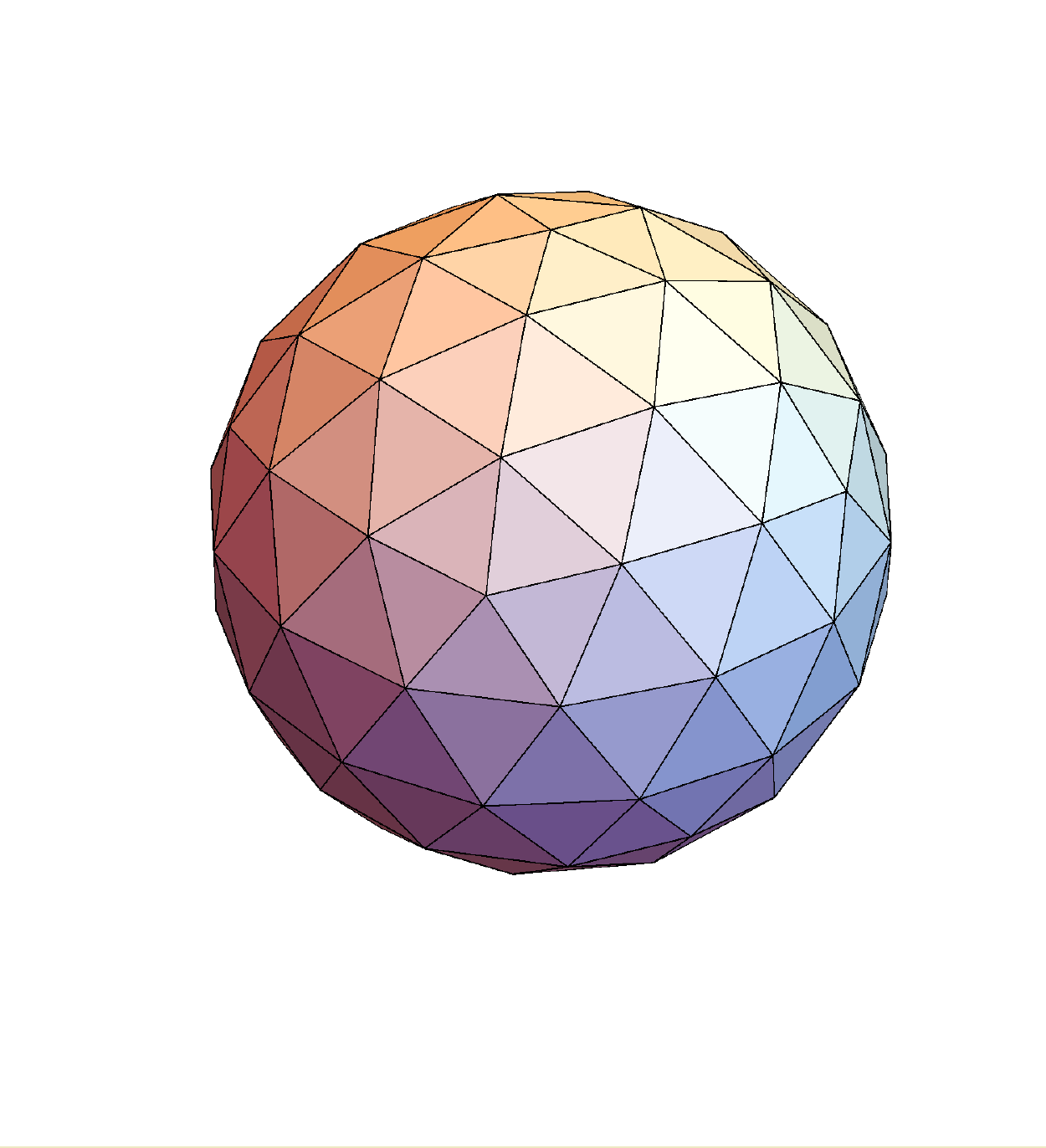}
\vskip -1.0 cm
\caption{\label{fig:icos} The  $s=3$ refinement of
the icosahedron with $N_s = 2 + 10s^2 = 92$ sites. The icosahedron on
the left is refined in the middle with $s^2 = 9$ equilateral triangles on 
each face and  then on the right vertices are projected  onto the unit
sphere. The resulting simplicial  complex preserves the icosahedral symmetries. }
\end{figure}

\paragraph{Finite Element Expansion: }
The second step is to expand the field, 
\be
\phi(x) \rightarrow \phi_{\sigma}(y) = E^0(y) \phi_0 +  E^1(y)
\phi_1 + \cdots +  E^{D} (y)  \phi _{D} \; , 
\ee
on each D-simplex $\sigma_D$ into  $D+1$  elements that
contribute exclusively to one site: $E^i(x_j) = \delta^i_j$. For
example, if we use  linear elements,
$E^i(y) = \xi^i$, in the interior and again eliminate, $\xi^0 = 1 - (\xi^1 + \cdots + \xi^D)$,
the field on the simplex  is 
\be
 \phi_{\sigma}(\xi) = \sum^D_{i =1} \xi^i (\phi_i - \phi_0) \; .
\ee
Using Eq.~\ref{eq:SimplicialAction} and taking the gradient, $\dd_i
\phi_{\sigma}(\xi) =  (\phi_i - \phi_0)$, the integrand for the
kinetic term in the action is a  constant trivially giving
\be
I_{\sigma} 
=  \frac{1}{2D!} \sum_{\<ij\>}  \sqrt{g}\; g^{ij} (\phi_i - \phi_0)
(\phi_j - \phi_0) 
\label{eq:vertex}
\ee
on each simplex. (The potential term, $m^2 \phi^2 + \lambda \phi^4$,
in Eq.~\ref{eq:SimplicialAction} is discussed in Sec.~\ref{sec:phi4}.) 
While this is the correct FEM kinetic term, one inconvenience is that
our arbitrary choice of
eliminating $\xi^0$   appears to  break the symmetry between the
$D+1$ sites.  There is a symmetric treatment which has an   appealing
geometrical form.  We introduce $D+1$ vierbeins  normal to the surface
opposite to each  of the $D+1$ vertices (Fig.~\ref{fig:simplex} ).  These are not linearly
independent because
\be
\vec \nabla (\xi^0 +\xi^1 + \cdots + \xi^D) = \vec e{\;}^0  + \vec
e{\;}^1 + \cdots + \vec e{\;}^D  = 0 \; ,
\label{eq:VierbeinZero}
\ee
but the  gradient  in this  over-complete
basis of tangent vectors is well defined,
\be
\vec \nabla \phi(y) =  \vec e{\;}^0\phi_0 +  \vec e{\;}^1
\phi_1 + \cdots +  \vec e{\;}^D  \phi _{D}  \; ,
\ee 
and re-evaluating the action 
we get a symmetric form,
\be
I_{\sigma} = \frac{1}{2 D!}  \sum^D_{i,j = 0} \sqrt{g}\; \vec e{\;}^i
\cdot \vec e{\;}^j \phi_i \phi_j = \frac{1}{2D!}   \sum_{\<i,j\>} \sqrt{g}\;  (-\vec e{\;}^i
\cdot \vec e{\;}^j) (\phi_i - \phi_j)^2  \; ,
\label{eq:ScalarFEM}
\ee
as a sum over all  $D(D+1)/2$ links on each simplex. This is exactly equivalent to Eq.~\ref{eq:vertex} using the identity
Eq.~\ref{eq:VierbeinZero}. More interestingly this identity
(\ref{eq:VierbeinZero})  also gives the Laplace operator in link form  on the RHS of Eq.~\ref{eq:ScalarFEM}. An alternative
ansatz for this operator was introduced in Ref. ~\cite{Christ:1982ci }  in terms of the dual simplex as
\be
I_{\sigma} =\frac{1}{2}  \sum_{\<i,j\>} V^D_{ij} \frac{(\phi_i
  -\phi_j)^2}{l^2_{ij}} \; ,
\label{eq:DualFEMscalar}
\ee
where $ V^D_{ij} = l_{ij} S_{ij} $  is the product of the
length of the link ($l_{ij}$) times the surface volume ($S_{ij} $)
of the dual polytope normal to the link  $\<i,j\>$.  That is  $V^D_{ij} =
l_{ij} \wedge *l_{ij}/D$, using the Hodge star of the discrete exterior
calculus~\footnote{
The use of the discrete exterior calculus for forms, dual vectors, etc  is a useful approach but the
careful implementation of this formalism is difficult in a short review.}.
This geometrical form (\ref{eq:DualFEMscalar}) suggested in
  Ref.~\cite{Christ:1982ci} is very appealing but
with the exception of $D = 2$  it is {\bf not equivalent to linear finite
  elements} (\ref{eq:ScalarFEM}) introduced in both the 
Regge and FEM literature.  For $D >2$, it may in fact be superior.
This is one of many instances where a direct application of linear
finite elements is either inadequate or at least may not be  optimal for
quantum field theory.

\subsection{Laplacian on Riemann Sphere}

We have tested the finite element construction of the Laplacian on the 2D Riemann
sphere, $\mathbb S^2$. The isometric embedding of $\mathbb S^2$  into $\mathbb
R^3$ is a sufrace with $\vec r \cdot \vec r = 1$.  For our simplicial lattice, we begin with an
icosahedron and subdivide each of the 20 triangular faces into
a regular grid of  $s \times s$ triangles as illustrated in
Fig.~\ref{fig:icos}. There are a total of $N_s = 2 + 10 s^2$
vertices, $E_s = 30 s^2$ edges and $F_s = 20 s^2$ faces satisfying
Euler's theorem for a spherical topology $F_s - E_s + N_s = 2$. 
\vskip -2 cm
\begin{figure}[h]
\centering
\includegraphics[width=0.45\textwidth]{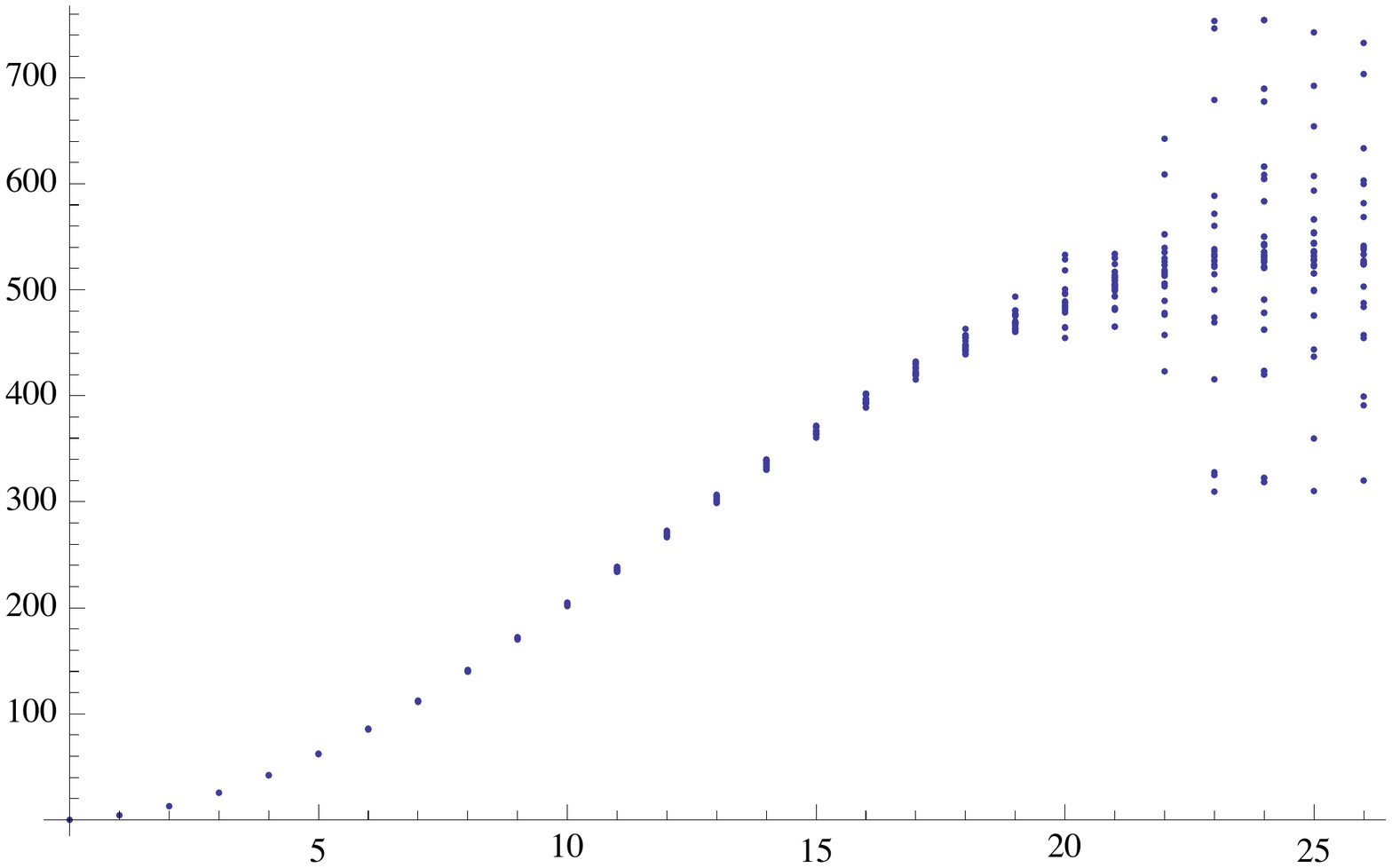}
\includegraphics[width=0.45\textwidth]{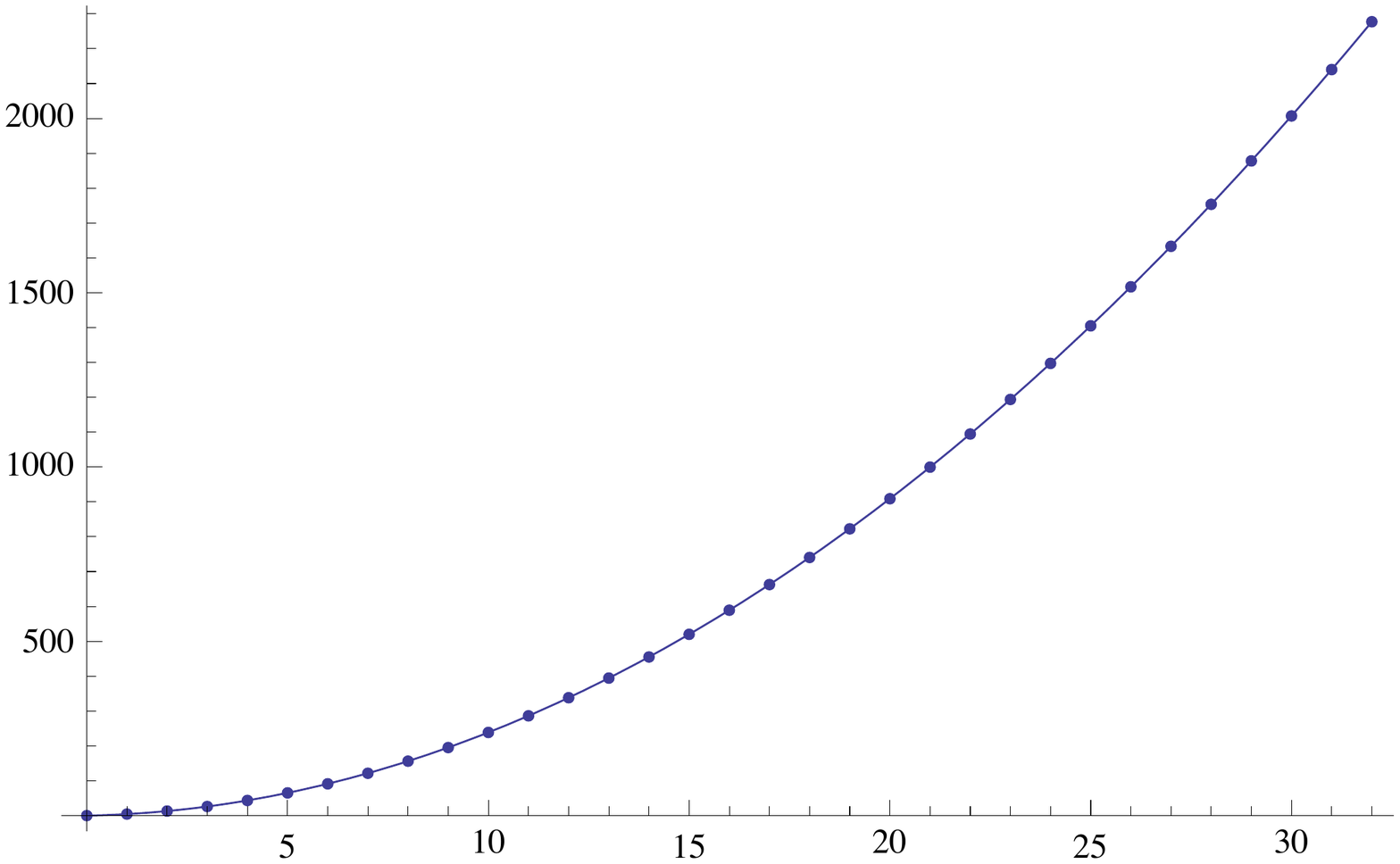}
\vskip -2 cm
\caption{\label{fig:FEMdiag} Left: The  $2l+1$  spectral  
 values for  $m \in [-l,l]$ are plotted against $l$ for $s
  = 8$.   Right: The averaged over $m$  fitted to $l + 1.00012 \; l^2 - 1.34281\times 10^{-7} \; l^3 - 0.57244
\times 10^{-7} \; l^4$   for $s =  128$   and $l \le 32$. }
\end{figure}
We then project radially the sites of the refined icosahedron onto the
unitsphere in $\mathbb R^3$ and set the lengths of the edges to the secant
distances, $l_{ij} = |\vec r_i - \vec r_j|$.  This defines our simplicial Riemann manifold  $({\cal M}_\sigma,g_\sigma)$ which is
a good approximation to the sphere that preserves the icosahedral
subgroup exactly, but the individual triangles are no longer regular.
The spectrum of the resultant FEM Laplacian matrix converges as
$O(s^{-2})$ to the continuum values as illustrated in Fig.~\ref{fig:FEMdiag}.

%%%%%%%%%%%%%%%%%%%%%%%%%%%%%%%%%%%%%%%%%%%%%%%%%%%%%%%%%%%%%%%%%%%%%%%%%%%%%%%%%%%%%%%%%%%%%%%%%%%%%%%%%%
\section{Interacting Scalars\label{sec:phi4}}

After testing the spectrum for the free scalar, there remains the
question of how to formulate  interacting field theories.  

As our first test, we drop the radial (time) direction in the 2+1
dimensional radially quantized $\phi^4$ theory.  In the continuum,
this example is equivalent to the stereographic projection of the
Ising (c=1/2 minimal model) CFT from ${\mathbb R}^2$ to ${\mathbb S}^2$.
%\be
%S=\int d^D x \, \sqrt{g} \left[\half g^{\mu \nu} \dd_{\mu} \phi \dd_{\nu} \phi + \lambda (\phi^2-\frac{\mu^2}{2\lambda})^2\right]
%\ee  
Since this map on the Riemann sphere  is conformal up to a Weyl rescaling, all correlation
functions are computable analytically at the fixed point, providing an
ideal rigorous test of convergence to the continuum.
 
Our  FEM partition function is given by
\be
Z=\int [D\phi] \text{exp}\left[-\sum_{\<ij\>}\frac{V^D_{ij}}{l^2_{ij}}
  (\phi_i - \phi_j)^2 - \sum_i V^D_i \lambda (\phi^2_i -
  \mu^2/2\lambda)^2\right] \; ,
\label{eq:FEMaction}
\ee
where for the kinetic term we used  weights, $V^D_{ij}$,
in Eq.~\ref{eq:DualFEMscalar}. For the potential term, we use
a  local  approximation to FEM integral,
as opposed the weighted average of the mass term discussed 
in Sec.~\ref{subsec:Mass}. The local weight, $V^D_i$, is  the  dual  Delaunay
volume  at $i$.  This  approximation is equivalent to dropping irrelevant  higher dimensional operators in a
renormalizable quantum field theory. Monte Carlo simulations for this
action are extremely efficient using the Brower-Tamayo form~\cite{Brower:1989mt}  of the
Wolff Cluster algorithm~\cite{Wolff:1988uh}. 

\begin{figure}[h]
\centering
\includegraphics[width=0.50\textwidth]{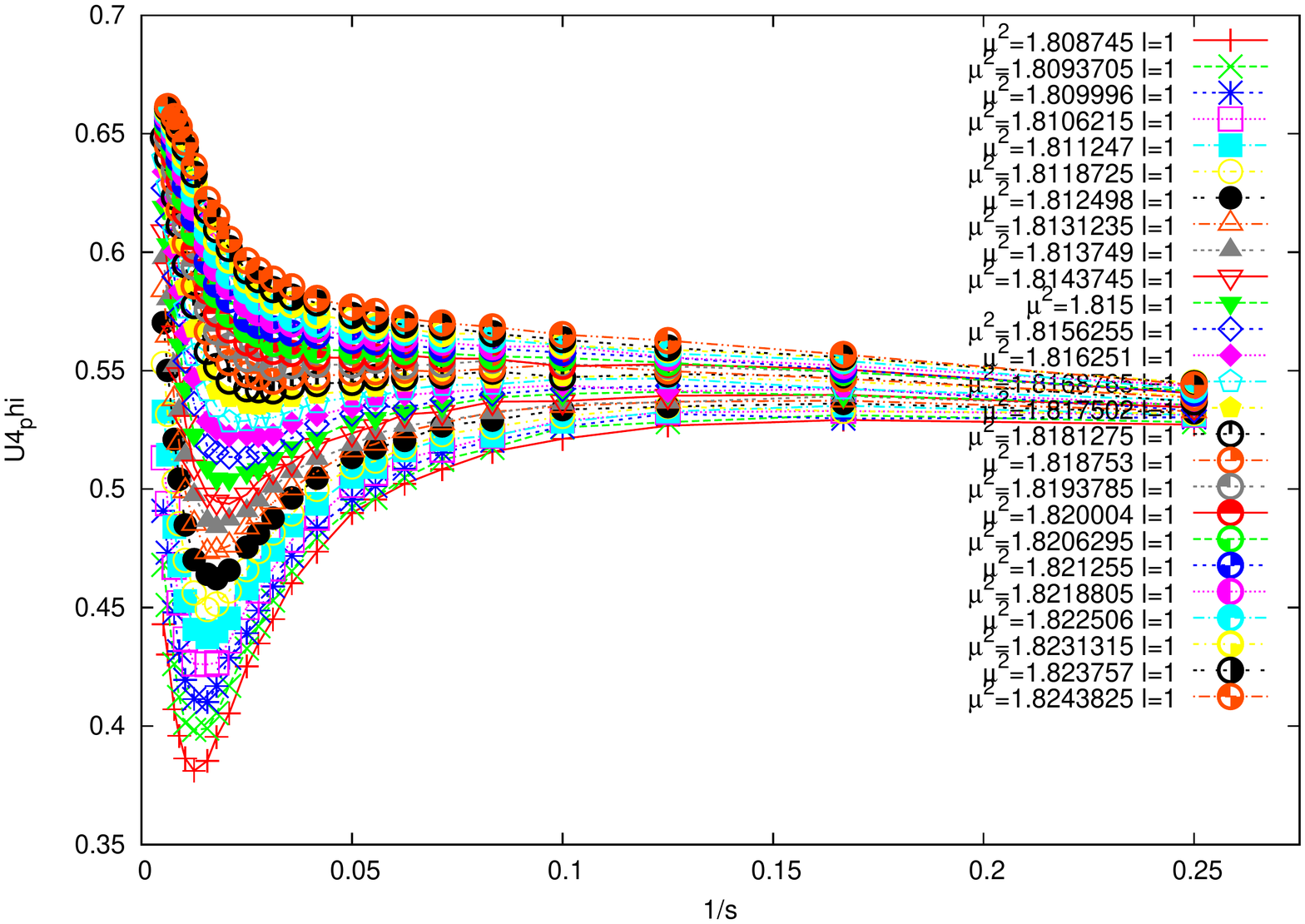}
\includegraphics[ width=0.40\textwidth] {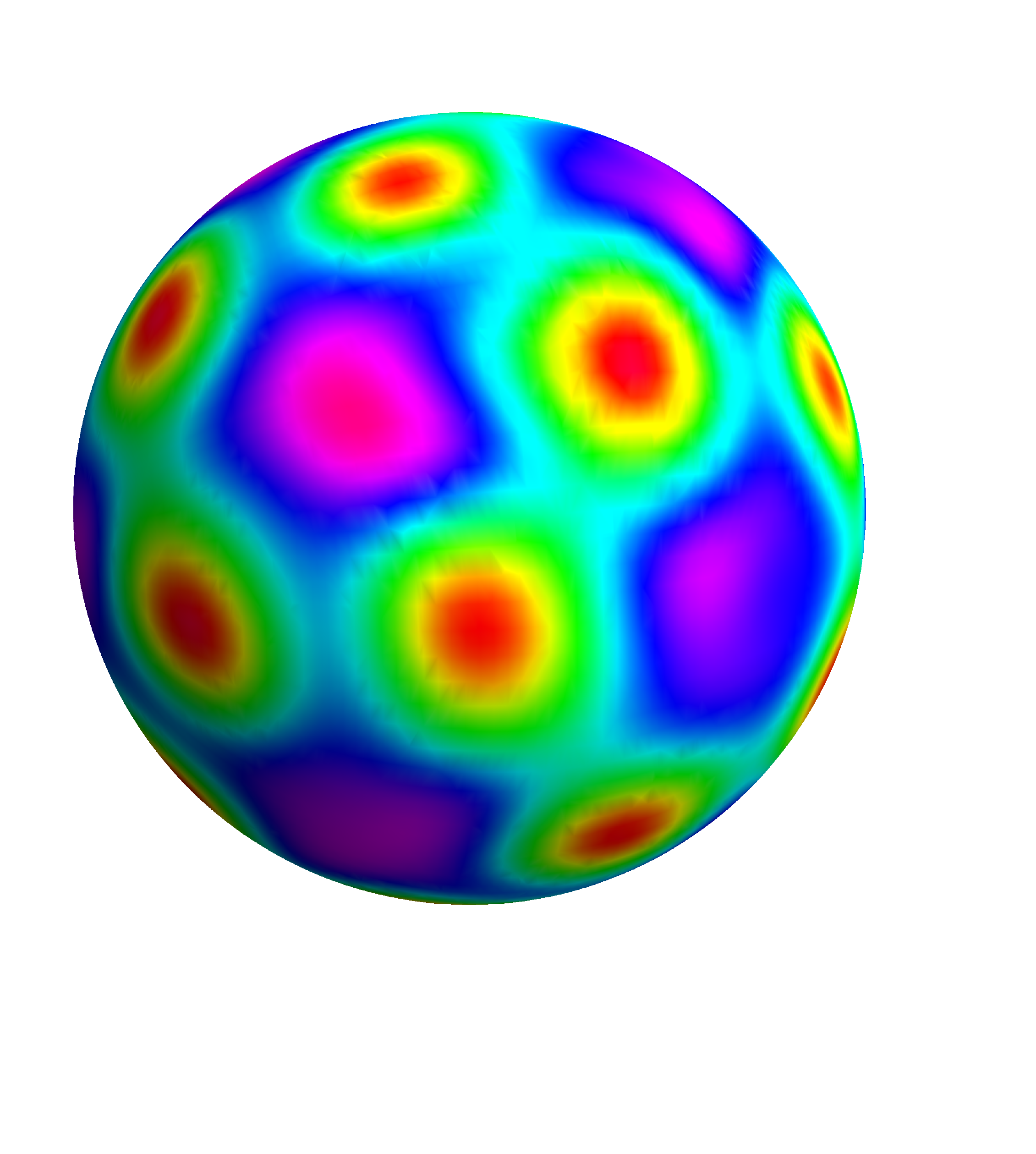}
 \caption{\label{fig:NoCT} On the left the Binder cumulant for the FEM
  Lagrangian with no QFE counter term. On the right the  amplitude of
  $\< \phi^2_i\>$  in simulations with the  unrenormalized FEM
  Lagrangian.}
 \end{figure}

 Our first  test of this was disastrous!  Evaluating  the Binder cumulant, 
\be
U_B(\lambda,\mu,s) = 1 - \frac{\<M^4\>}{3 \< M^2 \>\<  M^2\>}  \quad,
\quad M = \frac{1}{N_s} \sum_i \phi_i \; ,
\ee
we  attempted to  find the critical surface by tuning $\mu^2 \rightarrow  \mu^2_{cr}$ at fixed $\lambda =1$ in the continuum  limit, $a \sim 1/s
 \rightarrow  0$,  as illustrated on the left in Fig.~\ref{fig:NoCT}.  For moderate $s$, the Binder cumulant
 appears to stabilize close to the  analytical value~\cite{Deng:2003} of
 $U^*_B  = 0.567336(6)$, however as we  approached closer to the
 continuum at large $s$, it
completely destabilized. On the right in Fig.~\ref{fig:NoCT}, we
see the explanation. The  ensemble average, $\< \phi^2_i\>$, has large low mode
distortions. We see that the
regions  near the  original  12 poles of the icosahedron  and the region  at triangular faces of the icosahedron appear to  go critical at different values of
$\mu^2$. There is no critical surface with the  Wilson-Fisher fixed point. The conventional FEM
action (\ref{eq:FEMaction}) fails for  quantum field theory. 

%%%%%%%%%%%%%%%%%%%%%%%%%%%%%%%%%%%%%%%%%%%%%%%%%%%%%%%%%%%%%%%%%%%%%%%%%%%%%%%%%%%%%%%%%%%%%%%%%%%%%%%%%%
\subsection{Counter Terms for QFE}

\begin{figure}
\centering
\begin{minipage}{.5\textwidth}
  \centering
  \includegraphics[width=.8\linewidth]{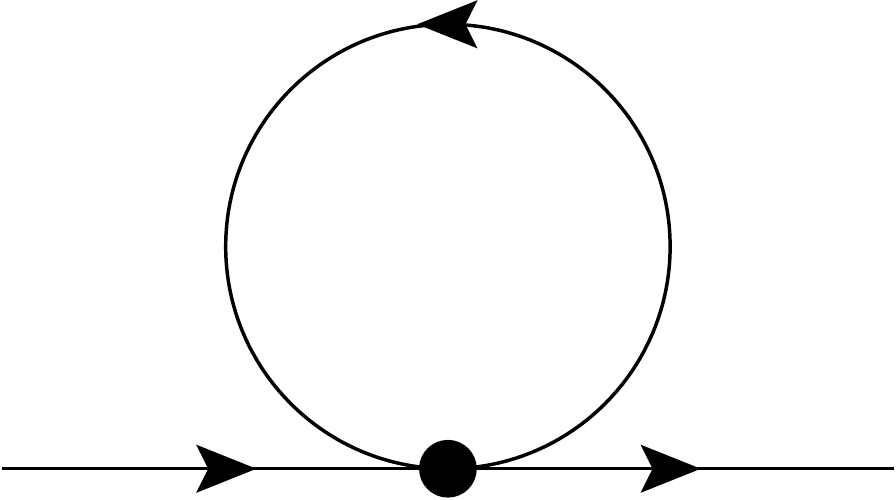}
\end{minipage}%
\begin{minipage}{.5\textwidth}
  \centering
  \includegraphics[width=.8\linewidth]{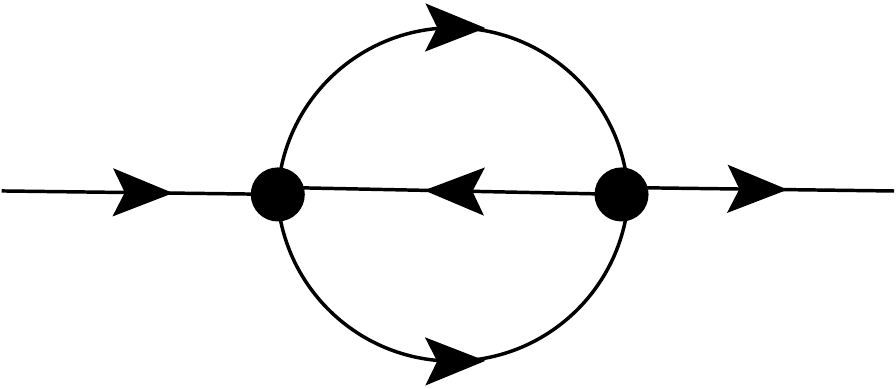}
\end{minipage}
\caption{\label{fig:loops} Left: The logarithmic divergence
  occurs at the one loop self energy diagram in 2D. Right: The two
  loop self energy diagram is finite in 2D but logarithmically
  divergent  in 3D. }
\end{figure}

The problem with FEM  for quantum field theory is fundamental. 
Although the proper use of FEM does guarantee convergence for  the classical equations for smooth solutions on any scale 
well separated from the lattice spacing (or inverse ultraviolet
cut-off $\Lambda  = 1/a \sim s$), the  quantum field path integral samples 
all scales with  divergent contributions in the ultraviolet (UV).  
Since the lattice is not quite regular, each lattice point has a 
slightly different UV cut-off $\Lambda_i = 1/a_i$. Fortunately in 
2D, the  $\phi^4$ theory is super-renormalizabie with the UV
divergence  occuring   only in the one-loop graph on the
left of Fig.~\ref{fig:loops}.   On the lattice this one loop diagram can be  
computed from our  free lattice propagator by inverting the kinetic term matrix,
 \be
K_{ij}(m^2_0) =\frac{V^D_{ij}}{l^2_{ij}} + m_0^2 V^D_i \delta_{ij} \; ,
\ee
choosing  $a^2 m^2_0=1.8/N_s$   to  fix the physical   mass scale
independent of  the cut-off.

\begin{figure}
\centering
\begin{minipage}{.6\textwidth}
  \centering
  \includegraphics[width=.9\linewidth]{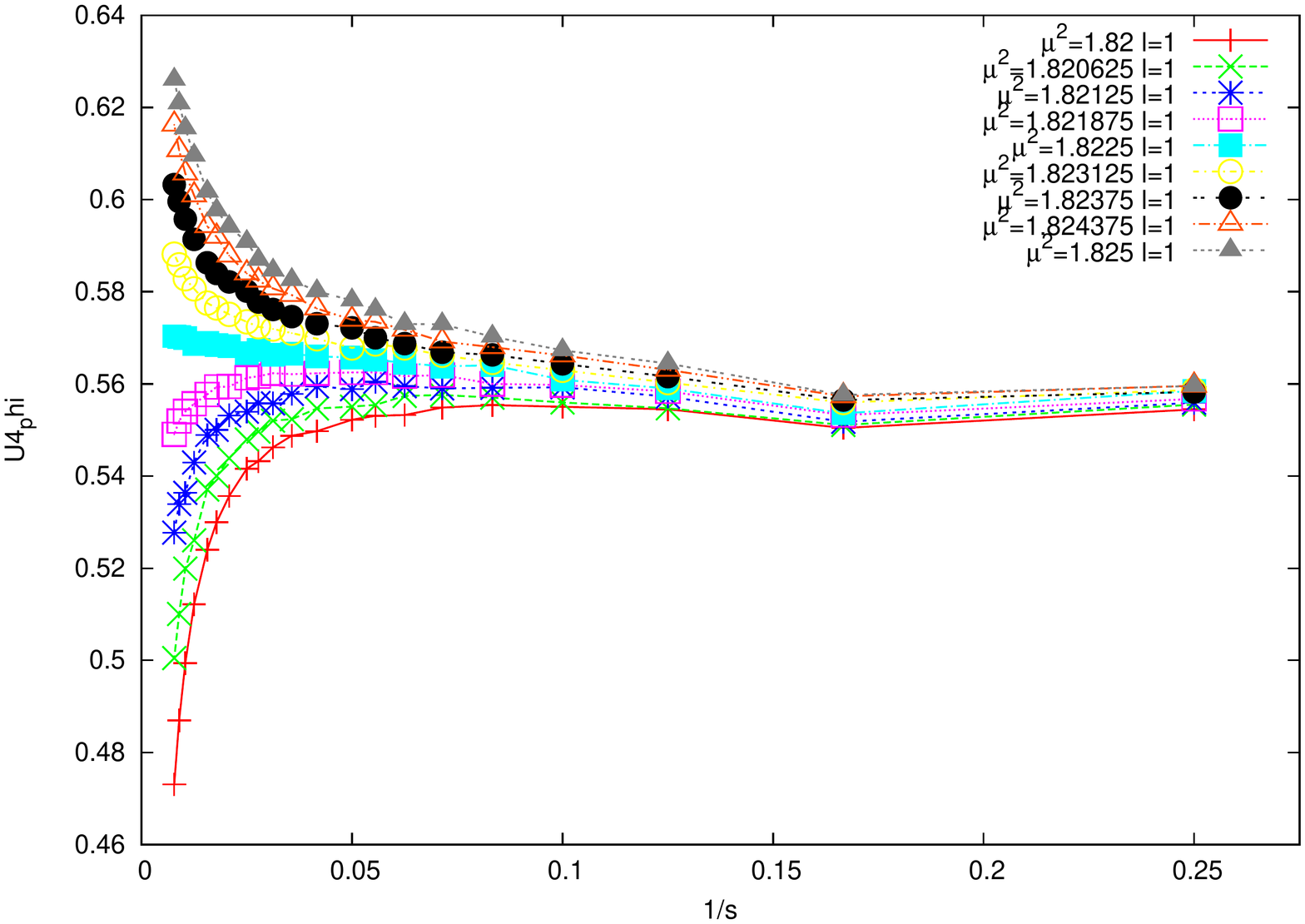}
\end{minipage}%
\begin{minipage}{.4\textwidth}
  \centering
  \includegraphics[width=.9\linewidth]{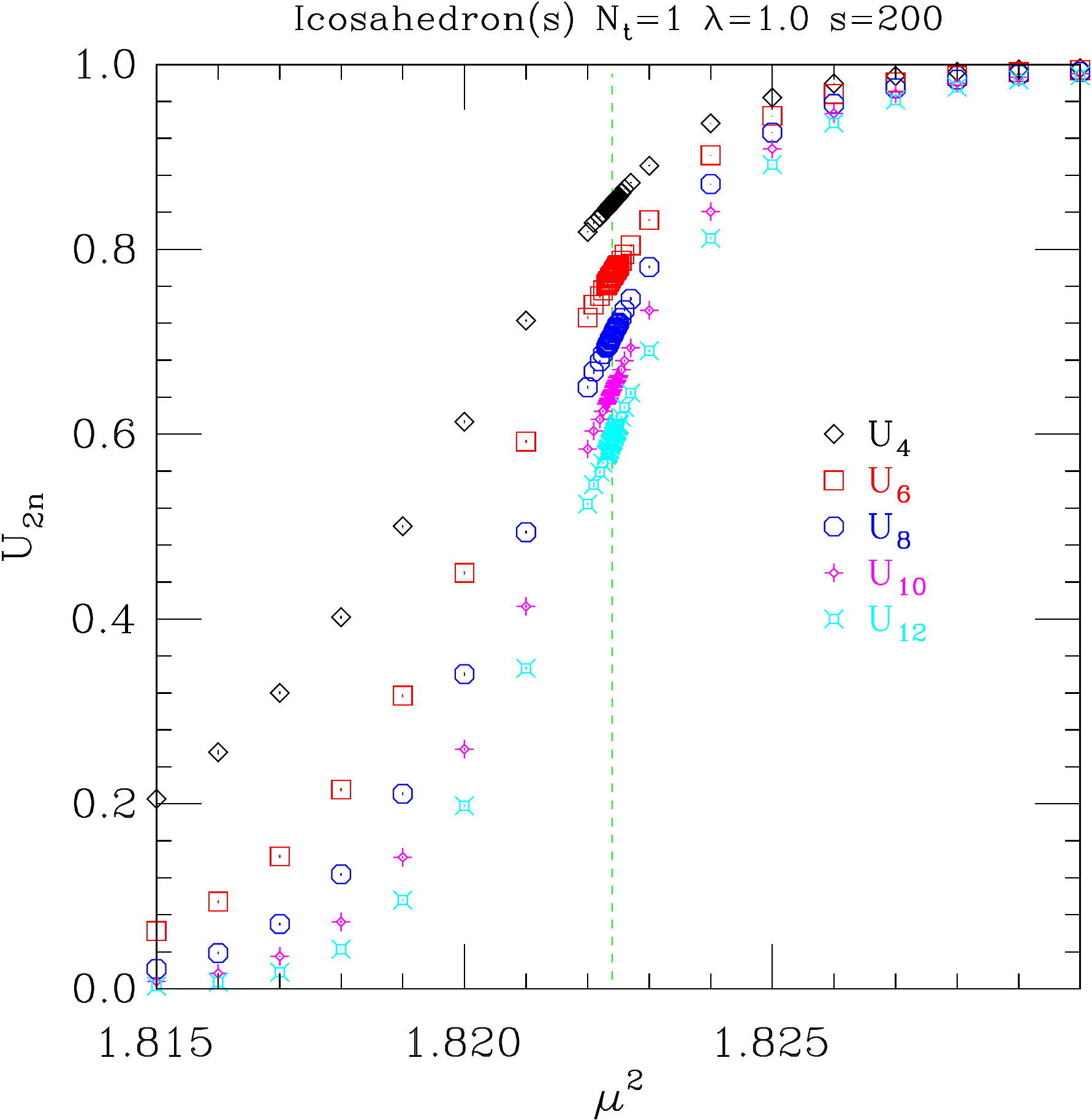}
\end{minipage}
\caption{\label{fig:binder} Introducing the counter term into the QFEM
  Lagrangian the Binder Cumulants approach fixed continuum values. On
  the left is the $U_B \equiv 2 U_4/3 =  1 - \< M^4\>/3 \<M^2\>^2$  and on the right
  a number of higher cumulants also approaching continuum as the
  cut-off is removed. }
\end{figure}

On the 2D Riemann sphere,  we find the one loop divergent term, 
\be 
\Delta m^2_i = 6 \lambda \left[K^{-1}\right]_{ii} \simeq \frac{\sqrt{3}}{8 \pi} \lambda
\log(1/m^2_0 a^2_i)  =  \frac{\sqrt{3}}{8 \pi} \lambda 
\log(N_s ) + \frac{\sqrt{3}}{8 \pi} \lambda \log(a^2/a^2_i)
\ee 
depends on an effective position dependent lattice spacing $a_i$.  This surprising result deserves a
comment. As expected, the graph is logarithmically divergent in the cut-off
$\Lambda^2 = 1/m^2 \simeq s^2$, however fortunately the FEM
prescription has removed the spatial dependence for the overall coefficient of
the log divergence in the $\log(N_s) $ term.  The ``charge'' is not
renormalized and is given by  its exact
continuum result. Thus  to cancel the UV spatial dependence in FEM
quantum Lagrangian, we can subtract a finite counter term: $ \mu^2 \rightarrow \mu^2 +\delta \mu^2_i$, 
\be \delta \mu^2_i = -
6 \lambda \large( \left[K^{-1}\right]_{ii} -
\frac{1}{N_s}\sum_{j=1}^{N_s} \left[K^{-1}\right]_{jj} \large) \ee
Indeed this finite term is almost precisely given by the  log of the
conformal scaling that mapped the flat faces of icosahedron onto the
sphere. With this correction we arrive at our {\bf QFE} simplicial Lagrangian,
which now approaches the correct continuum CFT governed by the
universal Wilson-Fisher fixed point in the continuum as illustrated in
Fig.~\ref{fig:binder}. This is an essential and significant success. 
The resultant {\bf QFE Lagrangian} must include this quantum effect.
 In a future publication, we will report more generally on this
critical step of determining the counter term. The 3D $\phi^4$ theory
which  is also
super-renormalizable, has 2 divergent diagrams in
Fig.~\ref{fig:loops} remormalizing the bare mass term. A very similar construction
of counter terms appears to apply to the 3D radial $\phi^4$ theory which
we are beginning to test numerically. Other approaches to
counter terms are under investigation. For example the Pauli-Villars
approach appears to provide a more general
method for renormalizable field theories with the exception of
non-Abelian gauge theories.

%%%%%%%%%%%%%%%%%%%%%%%%%%%%%%%%%%%%%%%%%%%%%%%%%%%%%%%%%%%%%%%%%%%%%%%%%%%%%%%%%%%%%%%%%%%%%%%%%%%%%%%%%%
\paragraph{Binder Cumulants of QFE Theory:}
We have tested our QFE 2D theory on a variety of  exact correlation
functions for this c=1/2 CFT.  Here we report one such test, the Binder cumulants,
 to support our claim that the counter term has solved the problem of
reaching the correct Wilson-Fisher fixed point for the c = 1/2 CFT. In
particular high precision simulations up to
$N_s = 2 + 10 (800)^2 \simeq 6 \times 10^6$ sites on the Riemann sphere
determined the critical surface at $\lambda = 1$ to be
$\mu^2_{cr} = 1.822400(4)$ providing an evaluation of the Binder
cumulant to be $U^*_4 = 0.8500(3)$ in statistical agreement with the analytic value
of $U^*_4 = 0.851003(8)$.
 
A series of higher order Binder cumulants ~\cite{Binder:1981zz} were
used to provide an effective way to determine the critical
surface.  

The Binder cumulants are moments of the magnetization
$m_n = \< M^n\>$ where $M = \sum_i\phi_i/N_s$. The lowest Binder cumulant,
$U_B = 1-m_4/(3m_2^2) \equiv (2/3) U_4 $ and higher cumulants are in general linear
combinations of scale invariant magnetization ratios.

\begin{eqnarray}
U_6 & = & \frac{15}{8} \left( 1 + \frac{m_6}{30\ m_2^3} - \frac{m_4}{2\ m_2^2} \right) \nonumber \\
U_8 & = & \frac{315}{136} \left( 1 - \frac{m_8}{630\ m_2^4} + \frac{2\ m_6}{45\ m_2^3} + \frac{m_4^2}{18\ m_2^4} - \frac{2\ m_4}{3\ m_2^2} \right) \nonumber \\
U_{10} & = & \frac{2835}{992} \left( 1 + \frac{m_{10}}{22680\ m_2^5} - \frac{m_8}{504\ m_2^4} - \frac{m_6\ m_4}{108 m_2^5} 
  + \frac{m_6}{18\ m_2^3} + \frac{5\ m_4^2}{36\ m_2^4} - \frac{5\ m_4}{6\ m_2^2} \right) \\
U_{12} & = & \frac{155925}{44224} \left( 1 - \frac{m_{12}}{1247400\ m_2^6} + \frac{m_{10}}{18900\ m_2^5} + \frac{m_8\ m_4}{2520\ m_2^6}
  - \frac{m_8}{420\ m_2^4} \right. \nonumber  \\
  & & \qquad \left. + \frac{m_6^2}{2700 m_2^6} - \frac{m_6 m_4}{45\ m_2^5} + \frac{m_6}{15\ m_2^3} - \frac{m_4^3}{108\ m_2^6}
  + \frac{m_4^2}{4\ m_2^4} - \frac{m_4}{m_2^2} \right) \nonumber
\end{eqnarray}

As we approach the critical surface, we parameterized the Binder
cumulant with a truncated expansion around the critical point,
\be
U_{2n}(\mu^2,\lambda,s) = U_{2n,\text{cr}}+
a_{2n}(\lambda)(\mu^2-\mu_{cr}^2(\lambda)) s^{1/\nu}+ b_{2n}(\lambda)
s^{- \omega} \; .
\ee
We fixed $\lambda=1$ and use the known critical exponents $\nu=1$
and $\omega=2$.  Then we  tune $\mu^2$ to find the critical surface as
can be seen in Fig.~\ref{fig:binder}.  The explicit fitting is done
via simultaneous fits to all cumulants with common $\mu^2_{cr}$.
Because this is a truncated expansion, one must be careful to only
select data such that higher order terms in the expansion can be
ignored.

 As seen in Fig.~\ref{fig:binder}, the value of
the Binder cumulant can change drastically even at $s\sim 50$.
Nevertheless, we are able to compute the first 5 Binder cumulants. 
These fits all have $\chi^2/dof \sim 1$, giving numerical predictions for the analytically computable
cumulants $\{U_{4,cr},U_{6,cr},U_{8,cr},U_{10,cr},U_{12,cr}\}$.
\begin{table}
\begin{center}
\begin{tabular}{|c||c|c|c|c|c|}
\hline $U_{4,cr}^{\text{Analytic}}$&$U_{4,cr}^{\text{Numeric}}$&$U_{6,cr}^{\text{Numeric}}$&$U_{8,cr}^{\text{Numeric}}$&$U_{10,cr}^{\text{Numeric}}$&$U_{12,cr}^{\text{Numeric}}$\\ \hline
 0.851003(8)&0.8500(3)&0.7724(4)&0.7072(6)&0.6483(8)&0.5944(8)\\
 \hline
\end{tabular}
\caption{\label{tab:bin}The first 5 Binder cumulants.  For $U_{4,cr}$ we can compare to the analytic result, but the higher cumulants are numerical predictions.}
\end{center}
\end{table}
Analytical values for higher Binder cumulants are not yet available
for comparison.

%%%%%%%%%%%%%%%%%%%%%%%%%%%%%%%%%%%%%%%%%%%%%%%%%%%%%%%%%%%%%%%%%%%%%%%%%%%%%%%%%%%%%%%%%%%%%%%%%%%%%%%%%%

\section{Dirac Fermions on Simplicial Lattice\label{sec:fermion}}

Further subtleties are encountered for the
discretization of Dirac and Gauge fields that carry spin. 
 The Dirac field on a Riemann manifold,
\be
S = \int d^Dx \sqrt{g}\; \bar \psi [ {\bf e}^\mu (\dd_\mu - i 
\boldsymbol{\omega}_\mu(x)) + m ]\psi(x) \; ,
\label{eq:dirac}
\ee
is particularly subtle since it requires two new structures to deal
with half integer spin: i.) The components of the tangent vector  ${\bf e}^\mu(x) = e^\mu_a (x)
\gamma^a$, where $e^\mu_a $ is  the inverse of  $e^a_\mu$  entering
into the metric.  ii.) The spin connection $\boldsymbol{\omega}_\mu(x) \equiv {\bf \omega}^{ab}_\mu(x) \sigma_{ab}/4 
$, where $\sigma_{ab} = i[\gamma_a,\gamma_b]/2$ are  the Lorentz
generators for the Dirac spinors.   The action is 
invariant under diffeomorphism or the choice of
co-ordinates for the manifold. In addition  there is a local ``gauge'' invariance
allowing an arbitrary  $O(D)$ rotation (or Euclidean Lorentz transformation) in the
tangent  planes: $y^a \rightarrow
\Lambda^a_b y^b$ at $x$.  This acts on the spinors as a  gauge invariance in
the {\tt Spin}(D) covering group. The  spin connection and the vierbeins are 
related by the tetrad postulate,
\be
\dd_\mu {\bf e}^\nu  + \Gamma^\nu_{\mu,\lambda}{\bf e}^\lambda = i
  [\boldsymbol{\omega}_\mu, {\bf e}^\nu]  \; .
\label{eq:TetrdadPost}
\ee

On a flat manifold the spin connection is a pure gauge and can be set
to one by choosing a global tangent plane but on a curved manifold this
is not possible. Indeed not all manifolds admit a spin
structure.~\footnote{Apparently a spin structure on a vector bundle E exists if
  and only if the second Stiefel-Whitney class   vanishes. 
%See \href{https://en.wikipedia.org/wiki/Spin\_structure}{https://en.wikipedia.org/wiki/Spin\_structure} 
} 
 The special subtlety of
placing the Dirac equation on a  simplicial lattice is to provide a lattice realization of this spinor
geometry.

Our construction  introduces a compact spin gauge link,
$\Omega_{ij}$ in  the spinor covering   group ${\tt Spin}(D)$ of the
Euclidean Lorentz group $O(D)$ 
and a  lattice ``vierbeins'' ${\bf e}^{(i)j} = e^{(i)j}_a  \gamma^a$
which are  tangent vectors contracted with the gamma matrix at  site $i$ on the
outgoing geodesic  from $i$ to $j$ as depicted in
Fig.~\ref{fig:Triangles}.  On the lattice we choose these  tangent
vectors  to have unit length: $ \vec e{\;}^{(i)j}\cdot  \vec
e{\;}^{(i)j}=1$.  The naive kinetic term in the lattice Dirac
action, 
\be
S_\sigma = \half \sum_{\<ij\>} \frac{V^D_{ij}}{l_{ij}} ( \bar \psi_i
 e^{(i)j}_a\gamma^a\Omega_{ij} \psi_j + \bar \psi_j
 e^{(j)i}_a\gamma^a\Omega_{ji} \psi_i ) + \cdots
\label{eq:diracOmega}
\ee 
is  anti-hermitian by virtue of $\Omega_{ji} = \Omega^\dag_{ij}  \;
,\; \Omega_{ji} {\bf e}^{(i)j} = - {\bf e}^{(j)i} \Omega_{ji}$.  A Wilson  term will be added to remove doublers. 
The Dirac operator   in Eq.~\ref{eq:diracOmega} is manifestly gauge invariant by virtue of  
\be
\psi_i \rightarrow   \Lambda_i \psi \quad,\quad
 \bar \psi_j \rightarrow \bar \psi_j  \Lambda^\dag _j \quad,\quad
{\bf e}^{(i)j} \rightarrow   \Lambda_i {\bf e}^{(i)j} \Lambda^\dag _i \quad,\quad
\Omega_{ij} \rightarrow   \Lambda_i \Omega_{ij} \Lambda^\dag _j  \; .
\ee 
 One can motivate this ansatz  by starting from flat space
 with $\Omega_{ij}=1$ and make an arbitrary spinor
rotations ($\Lambda_i$) to reorient the tangent vectors to derive
this parameterization  with the spin connection,  $\Omega_{ij} = \Lambda^\dag_i \Lambda_j$,
as a pure  gauge.  We should also note that, by generalizing the
ansatz in Ref~\cite{Friedberg:1985wr}, our
Fermion lattice action  is {\bf not} given by expanding the 
field $\psi(x)$  in linear FEM even for 2D in flat space. In general
linear elements fail to give ``verbeins''
$ e^{(i)j}_a  \gamma^a$ parallel to the links. However in 2D,  we
have been able to remedy this by inventing a new Dirac QFE basis
formed from 3 flat sub-triangles meeting at the dual
site (or circumcenter)  of each triangle. This construction reduces
exactly to our  action (\ref{eq:diracOmega}) for a flat 2D manifold as
will be demenstrated in Ref.~\cite{BrowerDirac}.

%%%%%%%%%%%%%%%%%%%%%%%%%%%%%%%%%%%%%%%%%%%%%%%%%%%%%%%%%%%%%%%%%%%%%%%%%%%%%%%%%%%%%%%%%%%%%%%%%%%%%%%%%%
\subsection{Spin Connection}

%\href{https://en.wikipedia.org/wiki/Spin\_structure}{https://en.wikipedia.org/wiki/Spin\_structure}

Now we must construct explicitly the lattice spin connection and
``vierbein'' on the simplicial manifold (${\cal M}_\sigma,g_\sigma$)
that conforms to  the target
continuum Riemann manifold (${\cal M},g$).  The first difficulty with our simplicial complex is the 
assumption that the interiors of each simplex
is flat, which implies curvature singularities at the
vertices.  In general to deal with the lack of a well defined tangent
plane at the vertices, in Regge Calculus the Fermions are usually
placed on the dual lattice sites~\cite{Hamber:2009mt} at the
circumcenter
of each simplex.  However,
this solution is troublesome for our goal of simplicial lattice field
theory. With gauge fields on links, matter fields (scalar and Dirac)
must be on sites to maintain gauge invariance.  Our solution is to construct the spin links $\Omega_{ij}$
assuming a new simplicial manifold without these curvature singularities prior to
introducing elements linear or higher order for the action. For
example on the sphere the solution is to remove the singular curvature at
the sites by replacing the links by geodesics (great circles in 2D)
and the simplices by spherical elements describe in Sec~\ref{sec:sfem}.  This
hybrid  approach is good to $O(a^2)$ as demonstrated
numerically  on $\mathbb S^2$. A more general  relaxation 
algorithm is also outlined below for a general simplicial complex that
should  converge on a sufficiently faithful approximation to a smooth Riemann
manifold.
\begin{figure}[ht]
		\begin{center}
		\vspace{10pt}
\includegraphics[width=0.48\textwidth,keepaspectratio]{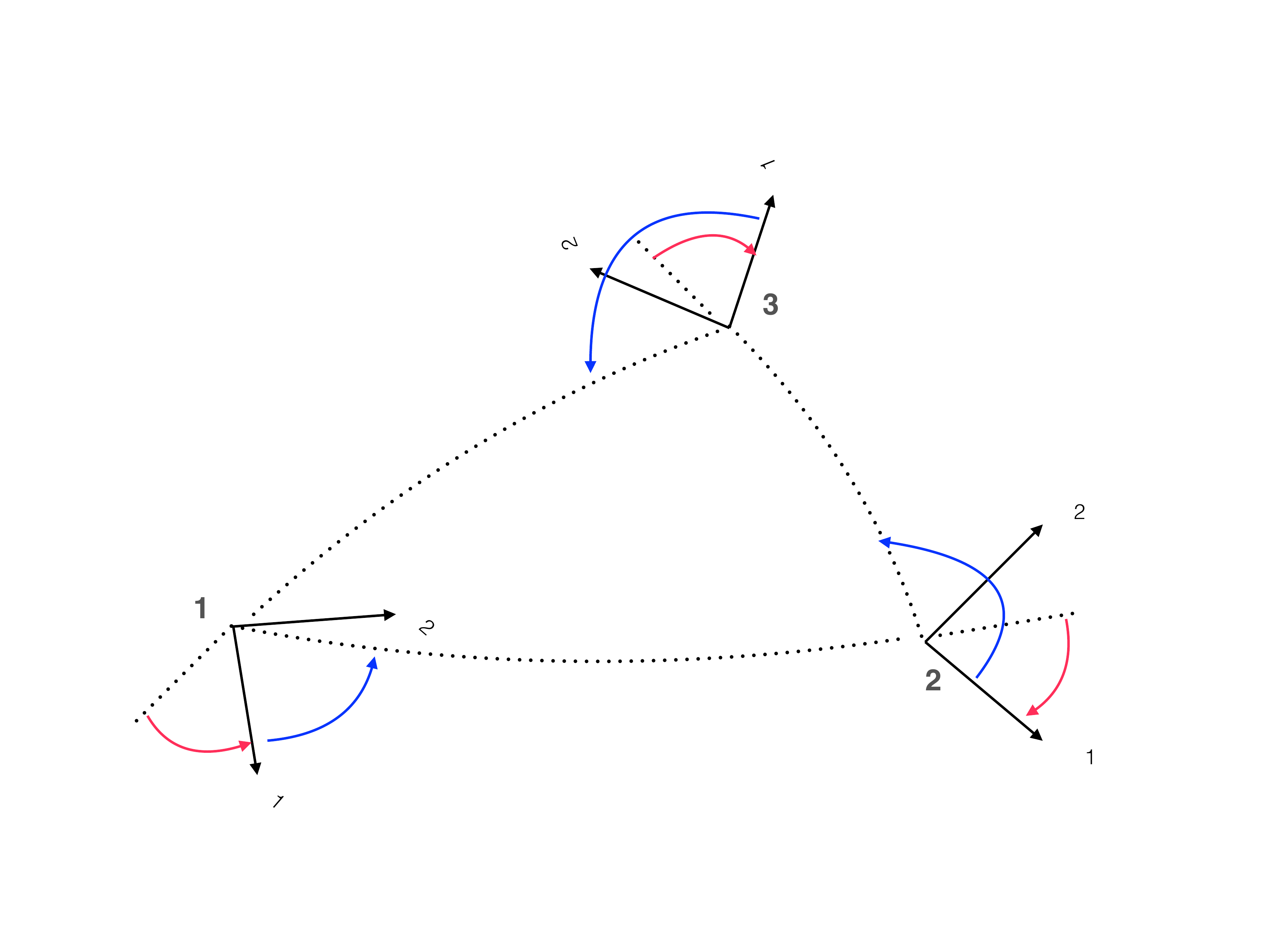}
\includegraphics[width=0.48\textwidth,keepaspectratio]{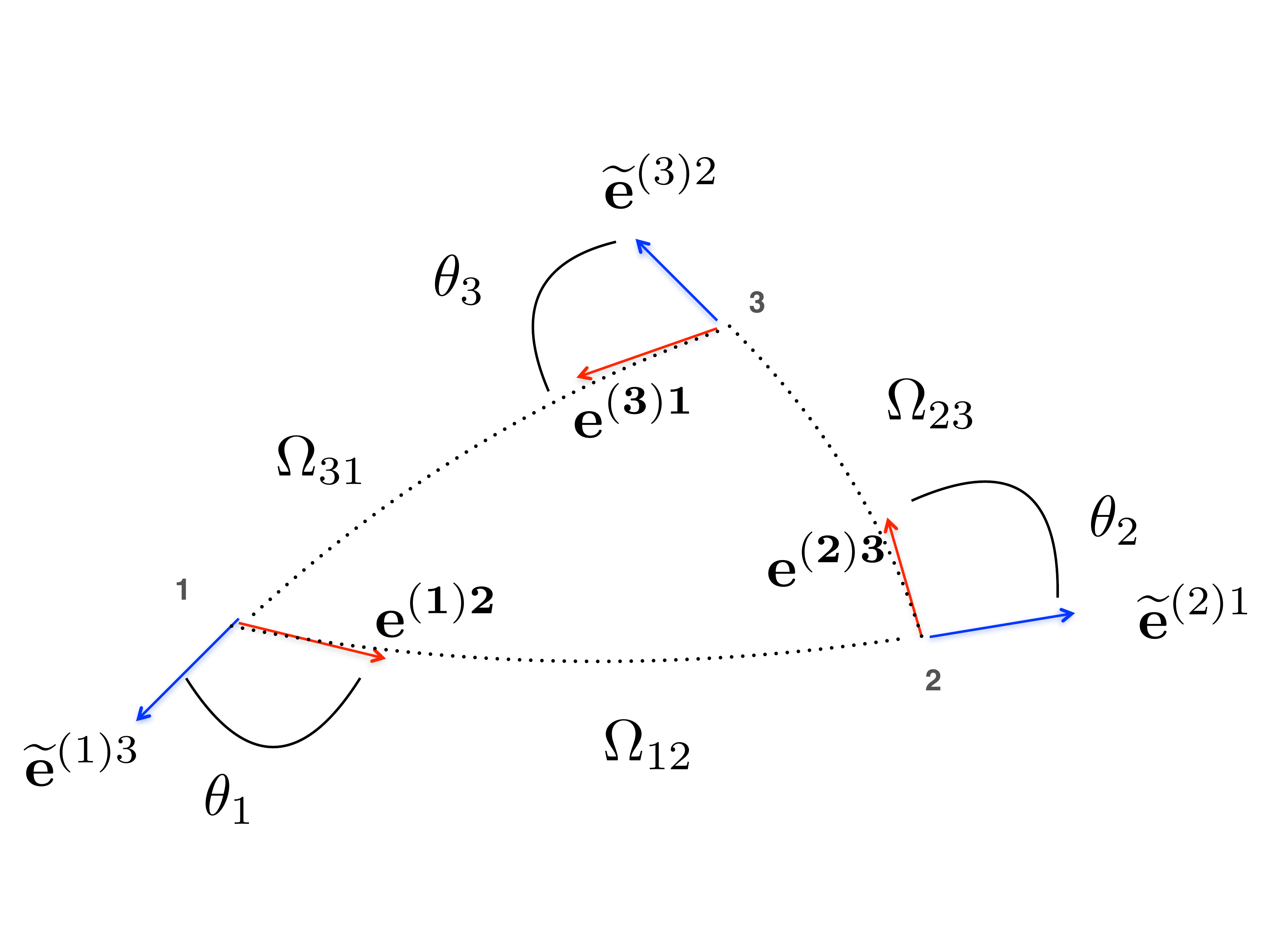}
   	\caption{\label{fig:Triangles} On the   left, vectors in the
          tangent plains and on the right the lattice spin connection,
          $\Omega_{12}$ and the outgoing and reflected  verbeins, ${\bf e}^{(i)j} =
          e^{(i)j}_a  \gamma^a$   and ${\bf \widetilde e}^{(i)j}
          \equiv - {\bf e}^{(i)j}$ respectively. }
   	\end{center}
\end{figure}

Naively,  $\Omega_{ij}$ simply describes the rotation
of a frame at (i) to a frame at (j).  However, the {\tt Spin}(D) group is a
\emph{double cover} of the Lorentz group $O(D)$.  This manifests itself
as a sign ambiguity in lifting $\Lambda_{ij} \in O(D)$ to the spinor representation, $\Omega_{ij}$. To fix
this ambiguity, we look at the integrated curvature invariant
associated with going around each triangular simplex. The algorithm
as illustrated  on the left in Fig.~\ref{fig:Triangles} proceeds  in 3 steps: {\bf i.)} 
Choose a random tangent frame at  $i$ and align  the  tangent vectors
$e^{(i)j}_a$  at site $i$ on  the geodesic from $i$ to all neighboring sites $j$. 
{\bf ii.)}  Parallel transport  the frame from  $i$ to $j$ along a  geodesic 
and  compute  the $O(D)$ Lorentz $\Lambda_{ij}$  rotation to the frame of
 $j$.  {\bf iii.)}  Map $\Lambda_{i j}
  \rightarrow s_{ij}\Omega_{ij}$  to the double cover in ${\tt Spin}(D)$ and
  choose the sign,  $s_{ij} = \pm 1$  to  minimize the
integrated curvature on the triangle: $\Omega_{\Delta}=\Omega_{ij}\Omega_{jk}\Omega_{ki}
  \hspace{5pt}\simeq  \hspace{5pt} 1 - O(A_{\triangle})$ consistent
  with  the area of the triangle, $A_\triangle$, vanishing in the  continuum limit.

  Choosing $s_{ij} = \pm 1$ is a global constraint. We can iterate
  fixing the signs, $s_{ij}$, on each triangle in order to satisfy
  $\Omega_{\Delta} \simeq 1$.  These signs are shared by adjacent
  triangles, so this procedure will either produce a \emph{specific}
  spin connection or fail to find a solution, if the manifold does not
  admit a spin structure.  For example on the sphere one can choose to
  expand the path one triangle at a time until we reach the last
  triangle. Now the total enclosed deficit angle on a sphere is
  $4 \pi$ steradians so the last small triangle has the right sign
  automatically satisfied.  Incidentally this follows essentially from
  Euler's theorem.  Higher genus 2D
  surfaces have $2^{2g}$ solutions.  For example for the 2 torus,
  ${\mathbb T}^2$, this procedure will lead to one of 4 different spin
  structures with period vs anti-periodic boundaries on the two
  non-contractible loops, referred to as Neveu-Schwarz/Neveu-Schwarz,
  Neveu-Schwarz/Ramond, Ramond/Neveu-Schwarz and Ramond/Ramond
  respectively in string theory.   In the language of category theory, the existence
  of a spin structure is a property shared between any simplicial complex
  and Riemann manifolds to which they  correspond.

The above procedure requires  calculating the  geodesic from
each site $i$ to its neighboring sites $j$ to  determine the angles
between the tangent vectors. For spherical geometry this is a straightforward
procedure because geodesics are known {\em a priori} to be  great
circles.   For more complicated surfaces, calculating geodesics
and performing parallel transport  can be
a difficult task and a more general method is  preferable. 
An alternative is to use a relaxation approach to 
match the lattice action to the continuum.  Here we sketch the
approach, with details of the implementation given in a future article.  We begin by 
introducing $\Omega_{ij}$ as free parameters to be fixed via a global
minimization, comparing the local curvature in $({\cal M},g)$ at each vertex $i$ to the
lattice approximation. To do  this we compare the 
continuum curvature at $i$ to the parallel transport around each adjacent
triangle $\triangle_{ijk}$:
\be
 S^{(i)}_\Delta \equiv e^{\textstyle i A^{\mu \nu}_{\Delta} {\boldsymbol R}_{\mu \nu}
   (i)} \quad \leftrightarrow \quad  \Omega^{(i)}_{\triangle_{ijk}} \equiv \Omega_{ij} \Omega_{jk} \Omega_{ki}
\ee
where the oriented area is defined by
$A^{\mu \nu}_{\Delta} = 
(1/6) [ (l^\mu_{12} l^\nu_{23} -  l^\nu_{12} l^\mu_{23 })+
\mbox{cyclic}]$ and the
\emph{local} curvature tensor by 
${\boldsymbol R}_{\mu \nu} (i) =i [\dd_\mu - i {\boldsymbol\omega}_\mu,
\dd_\nu - i {\boldsymbol \omega}_\nu]$ at site $i$. To
enforce this correspondence, one can simply minimize
a quadratic form, 
\be
\label{eq:spinmin}
R[\Omega_{ij}] = \sum_{\triangle, i}
Tr\left[(S^{(i)}_\Delta-\Omega^{(i)}_{\Delta})^\dag
  (S^{(i)}_\Delta-\Omega^{(i)}_{\Delta})\right] \; ,
\ee
by a procedure familiar to gauge fixing in lattice field theory. This
step has assumed  a particular  frame for the tangent planes,  so now we need
to find the  tangent vectors ${\bf e}^{(i)j}$ in that frame. To
do this, we recompute  the curvature as function of the  vectors ${\bf
  e}^{(i)j}$ on each triangle as illustrated Fig.~\ref{fig:Triangles}
on the right, assuming our values of $\Omega_{ij}$ and the  constraint:
$ {\bf e}^{(j)i} = -\Omega_{ji} {\bf e}^{(i)j}
\Omega_{ij} \equiv -  {\bf \widetilde e}^{(j)i} $. Parallel transport
anti-clockwise on a triangle has rotation matrices,
\be
 {\bf e}^{(i)i+1} =\Theta^{(i)} \; {\bf \widetilde e}^{(i)i-1}  =  e^{\textstyle  i \theta_i
 { n}^{(i)}_{ab} \sigma^{ab}/2 } \; {\bf \widetilde e}^{(i)i-1}   \quad,
\quad \Theta^{(i)}_{\Delta} = - \Theta^{(i)}  \Theta^{(j)}
\Theta^{(k)}  \; .
\ee
At each vertex,  the angle, $\theta_i$,  rotates ${\bf
    \widetilde e}^{(i)i-1}$ into ${\bf
e}^{(i)i+1}$,  as illustrated in Fig.~\ref{fig:Triangles}.
Thus  we can subsequently fix the ``vierbein'' by minimizing an
appropriate form such as
\be
E[{\bf e}^{(i)j}] =  \sum_{\triangle, i}
Tr\left[(S^{(i)}_\Delta-\Theta^{(i)}_{\Delta})^\dag
  (S^{(i)}_\Delta-\Theta^{(i)}_{\Delta})\right] \; , 
\ee
varying ${\bf e}^{(i)j}$ at fixed $\Omega_{ji}$. The result should determine ``vierbein''  on each tangent
plane up to an overall  irrelevant global rotation.  Numerical tests
and alternative algorithms for this  relaxation  procedure  are under
investigation.  

\subsection{Doublers and Chiral Symmetry}
\begin{figure}[ht]
		\begin{center}
		\vspace{10pt}
\includegraphics[width=0.4\textwidth,keepaspectratio]{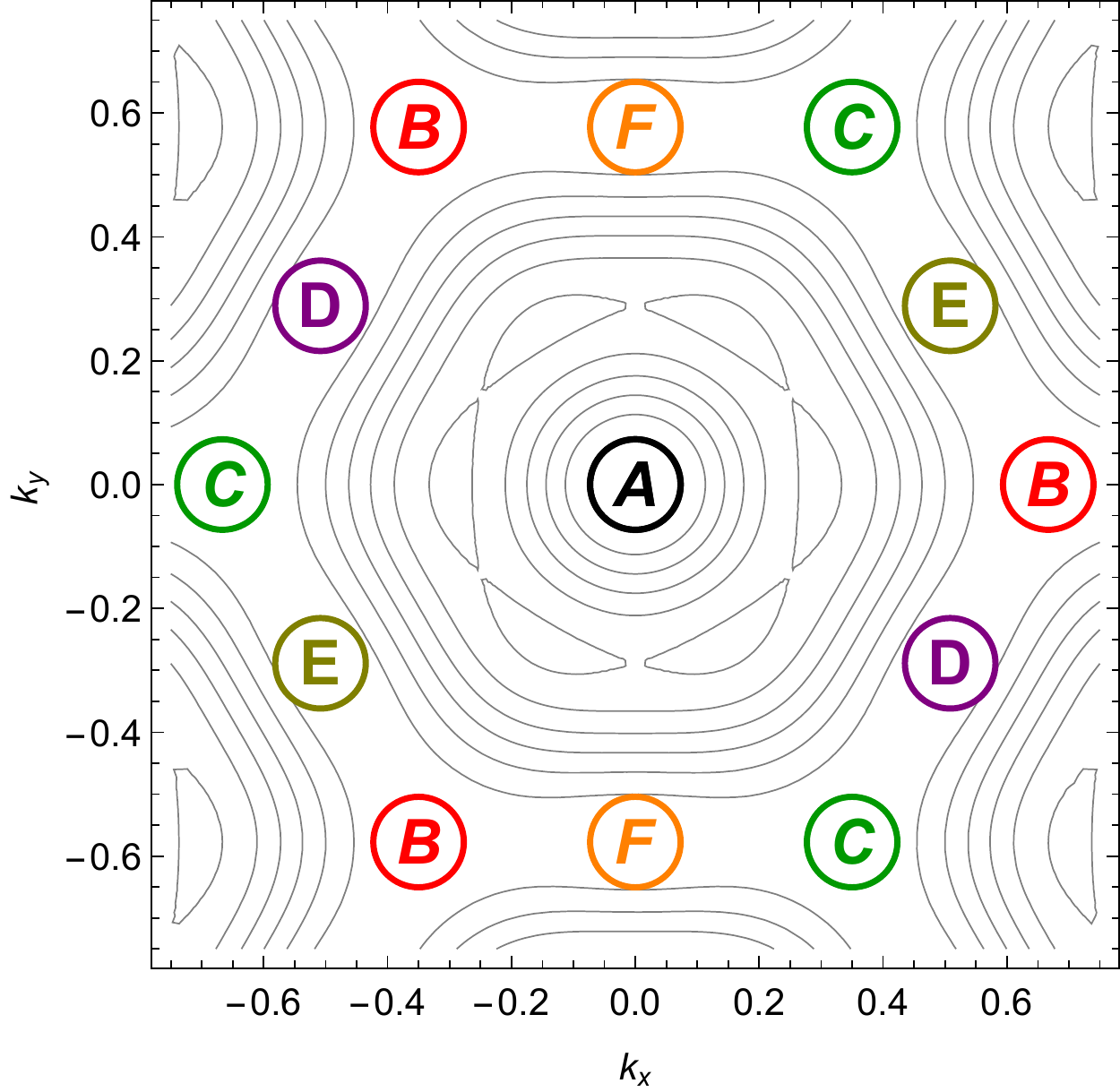}
\includegraphics[width=0.5\textwidth,keepaspectratio]{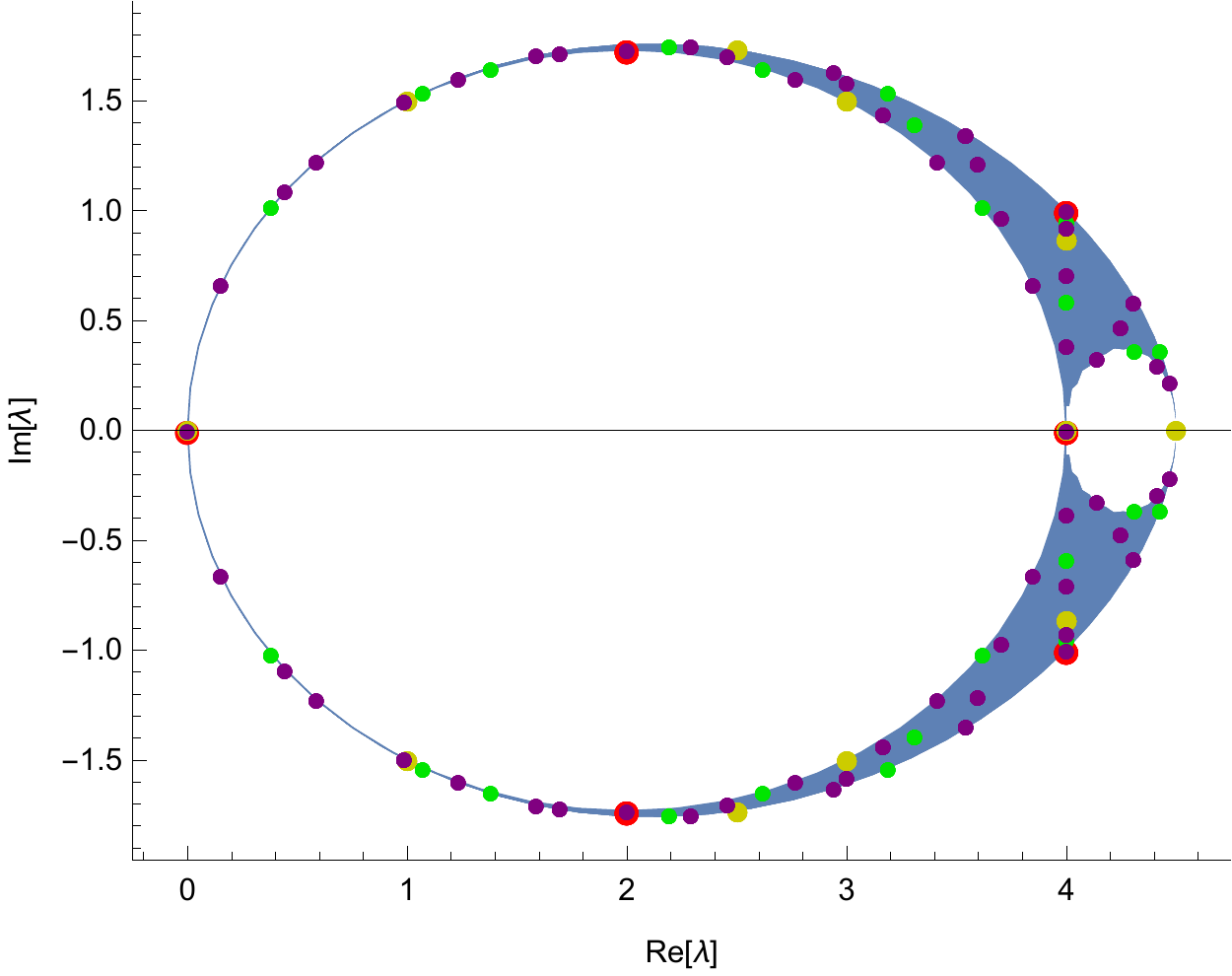}
   	\caption{\label{fig:flat} On the left the Brillouin zone for
          naive Dirac operator on a regular flat triangular lattice.
          The zero modes are labeled A-F. On the right, the infinite
          triangular lattice spectrum with the Wilson term (solid blue) compared to
          small lattices  with  16  (red), 36 sites (gold), 100
          (green) and 256  (purple) sites.}
   	\end{center}
\end{figure}

Next we must deal with the
usual issue of doublers and chiral symmetry for lattice fermions. 
In the lattice Dirac theory on a flat manifold, the doublers can be removed by
adding the Wilson term. In a  non-Abelian gauge
theory, the square of the Dirac operator, 
\be
[\gamma_\mu(\dd_\mu -i  A_\mu)]^2 = (\dd_\mu -i  A_\mu)^2 - \frac{1}{2}
\sigma^{\mu \nu} F_{\mu \nu} \; ,
\ee
is a sum of two terms which are referred as  the
Wilson term  and the clover term respectively, when placed on the
lattice. Similarly the square of the spinorial
Dirac operator on a curved manifold is  a linear combination of two
similar terms,
\be
 [{\bf e}^\mu_a  (\dd_\mu - i\boldsymbol{\omega}_\mu)]^2 =  \frac{1}{\sqrt{g}} \boldsymbol{D}_\mu
\sqrt{g} g^{\mu \nu} \boldsymbol{D} _\nu  - \frac{1}{2} \sigma^{ab} e_a^\mu e_b^\nu \boldsymbol{R}_{\mu \nu}
\ee
where $\boldsymbol{D} _\mu = \dd_\mu - i\boldsymbol{\omega}_\mu$. On
our simplicial complex, the first term will become our Wilson operator on a curved manifold,
\be \label{eq:wilsonterm}
S_{Wilson}= \frac{r}{2} \sum_{\braket{i,j}} \frac{aV_{ij}}{l_{ij}^2}
(\bar{\psi}_i-\bar{\psi}_j \Omega_{ji})(\psi_i - \Omega_{ij} \psi_j)
\; ,
\ee
and the second term is the lattice curvature in the spinor basis. Before utilizing this Wilson term
on curved manifolds, it is interesting to see its effect on
a flat  2D triangular lattice.  As depicted in Fig.~\ref{fig:flat},
in absence of the Wilson term, the hexagonal Brillouin zone in 2D on a
triangular lattice actually has 6 copies of the 2 component spinor, but
when the Wilson term is added the doublers are removed and the
spectrum comes close to  the circular  complex spectrum  of a lattice overlap
operator. 
Not surprisingly the qualitative effects on $\mathbb S^2$ are very
similar. We should also note that this Wilson term can be used
to introduce  Domain Wall  Fermions on a simplicial complex with an extra flat
 dimension  to restore chiral symmetry.

%%%%%%%%%%%%%%%%%%%%%%%%%%%%%%%%%%%%%%%%%%%%%%%%%%%%%%%%%%%%%%%%%%%%%%%%%%%%%%%%%%%%%%%%%%%%%%%%%%%%%%%%%%
\subsection{Spectral Analysis \label{sec:spectrum}}

\begin{figure}[ht]
\centering
\includegraphics[width=0.4\textwidth]{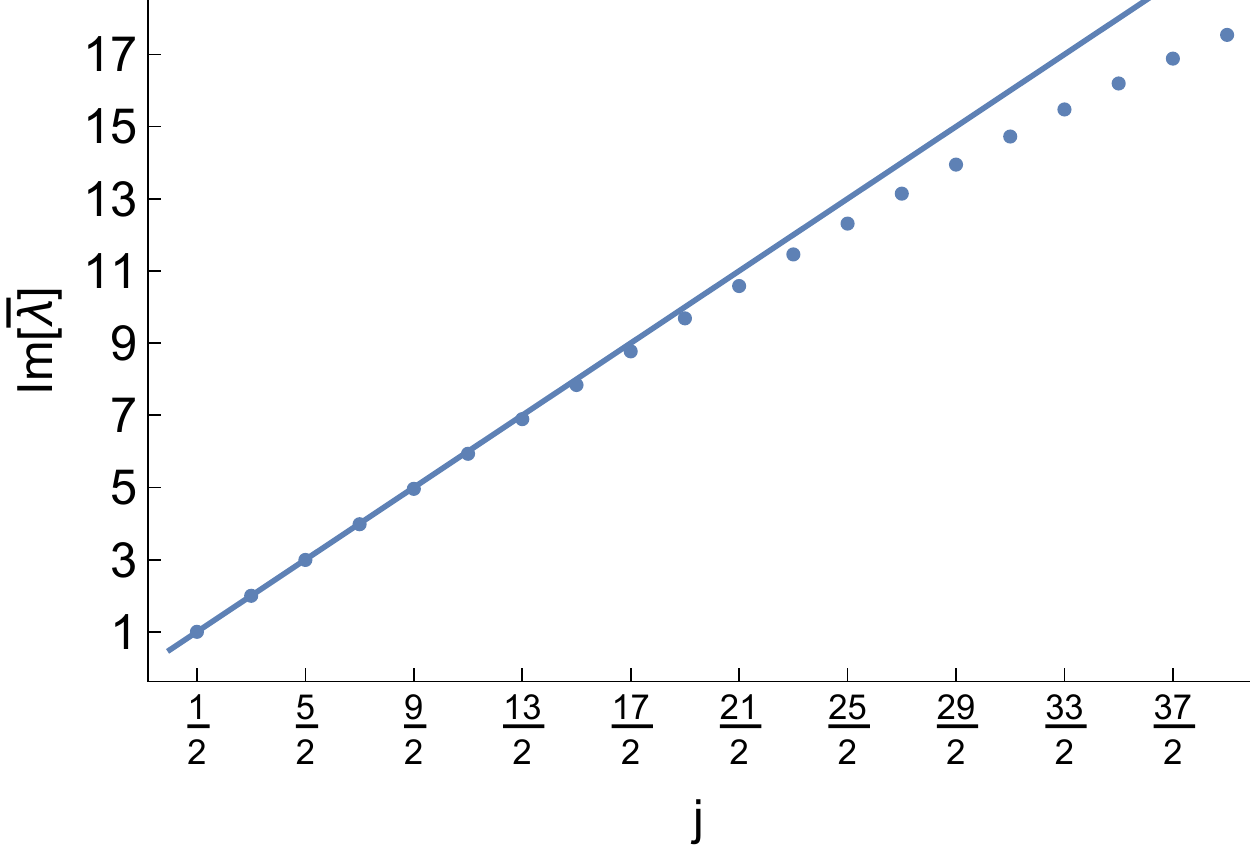}
\includegraphics[width=0.4\textwidth]{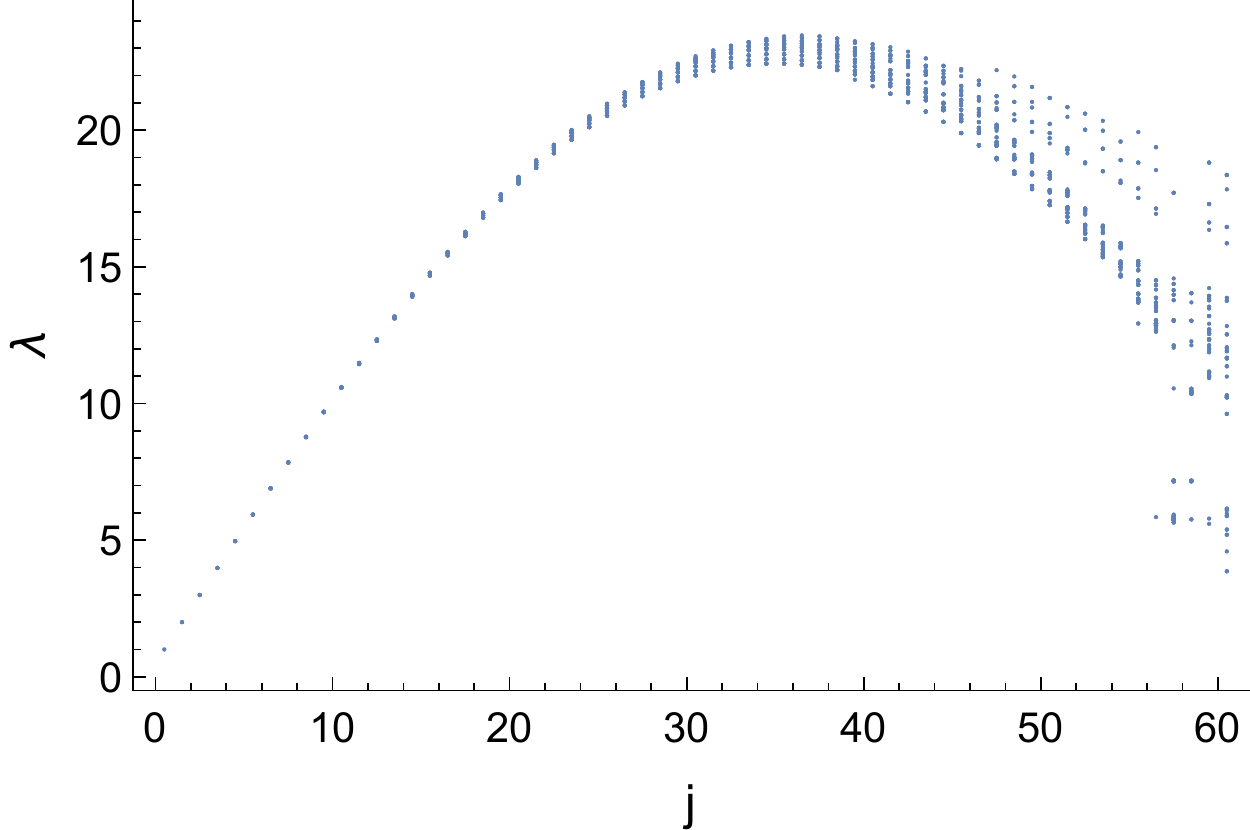}
\caption{\label{fig:FEMdiagfermion}  Left: The eigenvalue $\lambda$
  averaged over $m$  and  compared to the expected linear behavior. Right: The  imaginary part of the spectral eigenvalues  $\lambda_{continuum}=j+1/2$ for $m \in [-j,j]$ are plotted against $j = 1/2, 3/2,\cdots$ for $s
  = 24$. }
\end{figure}

 Numerical tests of the simplicial Dirac  framework on a $\mathbb S^2$ have been performed which demonstrate
convergence to the exact continuum 
spectra~\cite{Abrikosov:2002jr} as a function of the square
of the average lattice spacing,  $a^2 \sim 1/s^2$, in
Fig.~\ref{fig:FEMdiagfermion}.

\begin{figure}[ht]
  \begin{center}
    \includegraphics[width=0.7\textwidth]{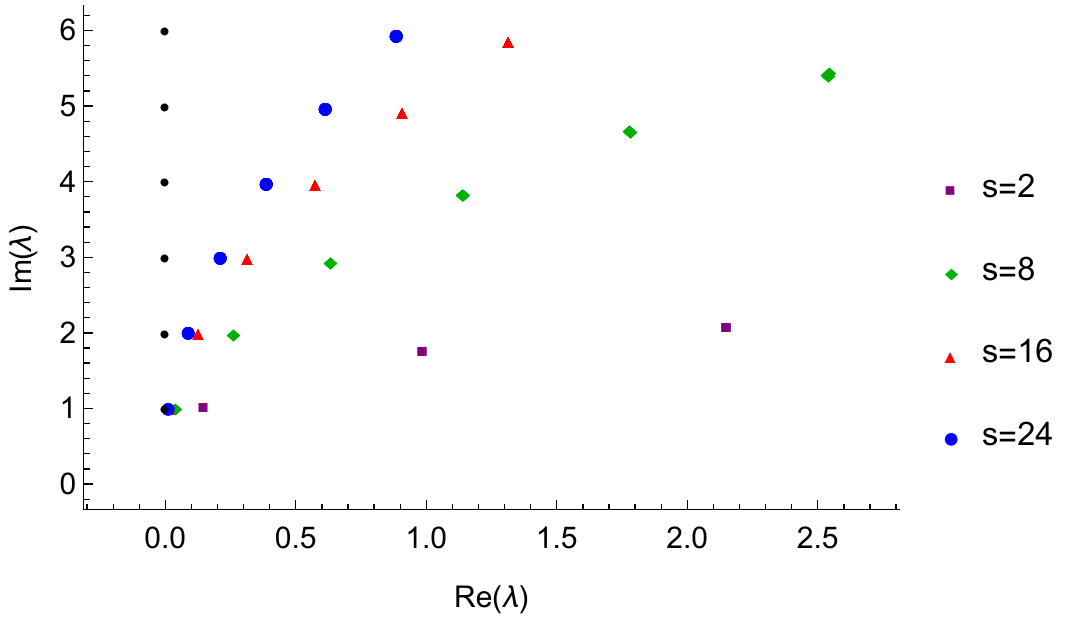}
  \end{center}
\vskip -0.5cm
  \caption{\label{fig:spectrum}Small norm eigenvalues for the full
    Wilson-Dirac spectra for various refinements of the icosahedron.
    As $s\rightarrow \infty$ the small eigenvalues approach the
    continuum limit $\lambda =  \pm i (j + 1/2)$ for .}
\end{figure}

The continuum Dirac operator, $\bf {D}$,  on $\mathbb S^2$ is anti-hermitian with $``\gamma_5"$
Hermiticity: $\sigma_3 \bf{D} = \bf{D}^\dag \sigma_3$ for D = 2. The
spectrum is gapped, with eigenvalues $\lambda = \pm i(j +1/2)$
with $2j+1$ degeneracy for total angular momentum
$j = 1/2, 3/2,\cdots$. Chiral fermoins on a sphere have no zero mode. The size of the gap scales inversely to the radius of the
$\mathbb{S}^2$.  The Euclidian continuum theory also admits a
parity/time-reversal symmetry ($\sigma_1, \sigma_2$ respectively) 
with the product being  charge-conjugation symmetry.

We can now build a $2 N_s\times 2 N_s$ Dirac-Wilson operator described
by Eq. \ref{eq:dirac} and analyse its spectrum.  The results for
$s=2,8,16,24$ can be seen in Fig.~\ref{fig:spectrum}.  The full
shape of the spectrum is reminiscent of the arc-like shape seen for
fermions on a flat triangular lattice.  As $s\rightarrow \infty$ the
arc gets steeper for small-norm eigenvalues approach the continuum
result as seen in Fig.~\ref{fig:spectrum}.  The real part of the
spectrum, which determines this arc's  shape, is determined by the
Wilson term.  As the Wilson term is turned off, the eigenvalues
collapse on the imaginary line but suffers from doublers. 

We find that the first 3 levels of eigenvalues are  exactly
degenerate---corresponding to the protected icosahedral symmetry.
Further, the $\{\sigma_1,\sigma_2,\sigma_3\}$ symmetries are
preserved on the lattice.  The lattice eigenvalues approach the continuum as
$1/s^2$.  For example, if we look at the imaginary part for
$\lambda=4 i$ (the lowest multiplet not protected by icosahedral symmtry), it approaches the
continuum as
$(3.99987\pm0.00031)-(11.35\pm0.57)/(s+(0.163\pm0.020))^2$.  The
Wilson term contains an extra factor of lattice spacing and thus
vanishes more slowly as
$(-0.00595 \pm 0.00042)+(9.576 \pm 0.018)/(s + (0.348 \pm 0.026))$ at
the fourth eigen-level. We have also checked that the eigenvectors of our lattice Dirac operator
converge to the analytic continuum wave function for
the Dirac equation on $\mathbb S^2$. This is accomplished by
a  rotation of the gauge from our random tangent planes to be
aligned with the continuum solution in the $\{\hat{\theta},\hat{\phi}\}$ frame ~\cite{BrowerDirac}.

%%%%%%%%%%%%%%%%%%%%%%%%%%%%%%%%%%%%%%%%%%%%%%%%%%%%%%%%%%%%%
\section{Spherical Finite Elements\label{sec:sfem}}
%motivation

As we mentioned in our application to the Dirac action on
$\mathbb S^2$, we  assumed a spherical simplicial lattice for
the purpose of fixing the spin connection.  This naturally leads us to
consider if spherical elements could be constructed as a FEM
basis. Although an explicit construction is unnecessary, alternative
discretization schemes are interesting in their own right as possible
improvement schemes designed to accelerate convergence to the continuum.

Here we show that on $\mathbb{S}^2$, it is indeed
possible to design a finite element using spherical triangles whose
boundaries are the geodesics of the sphere.  We compute the spherical
finite elements using a method which we call the ``renormalization
group approach,'' and argue that it may be applied to more general
manifolds and to theories which are not conformal.  We also mention a
second approach to computing the spherical elements based on the
Riemann mapping theorem which is mathematically elegant although
less general.

The finite element approach can be expressed as a decomposition
of  the continuum action into a sum over disjoint integration domains,
\be
S = \int_{\cal M} \mathcal{L}[\phi,\partial\phi] = \sum_{\sigma_D}\int_{\sigma_D}
\mathcal{L}[\phi,\partial\phi] \; .
\ee
For a scalar field, let us  also assume that the domains in D-dimensions are defined by
D+1 sites and  the fields are determined by the value of the 
fields on the sites. To achieve this we need to  establish unique
boundaries between the domains following a natural generalization of the linear element in 
flat space. We define the edges ($\sigma_1$) to be geodesics, a natural generalization
of straight lines in flat space. For D > 2, we ascend to higher dimensional
boundary elements:  minimal 2D surfaces  ($\sigma_2$) with fixed
edges, minimal 3D  volumes ($\sigma_3$)  with fixed surfaces, etc. Finally
we define the  elements to be the unique solution to 
Laplace's equation with the given boundary data.  This procedure
guarantees that the general solution is a linear  function of 
the fields on the $D+1$ sites
\be
\phi(x) = \sum^D_{i = 0} E^i(x) \phi_i \; .
\ee
The elements $E^i(x)$ are therefore  solutions with $\phi_i=1$ and
$\phi_{j\neq i} = 0$ and the identity,
\be
 E^0(x) + E^1(x) + \cdots + E^D(x) = 1 \; ,
\ee
is guaranteed so the constant field is preserved.  By design this prescription in flat space  
reduces to  the linear FEM,  $E^i(x) = \xi^i$, in terms of  barycentric
coordinates. 

Once the weight functions are known, the integral over the functional
of the elements  defines the discrete massless kinetic term
or Laplacian operator on each element,
\be
S_{\Delta} = \frac{1}{2} \sum_{\<i,j\>}K_{i,j}^\Delta  (\phi_i -
\phi_j)^2 \quad \mbox{where} \quad
K_{i,j}^\Delta = - \int_{\Delta} d^2 x \sqrt{g(x)} g^{\mu
  \nu}(x)\partial_\mu E^i(x) \partial_\nu E^j(x) \; .
\ee
We now we restrict our explicit construction to $\mathbb S^2$, although
generalizations to other smooth  Riemann manifolds should follow
a similar approach. 

\paragraph{Renormalization Group Approach:}
Our task is to solve a classical field equation on a curved manifold
with prescribed  boundary values.  This is exactly the kind of problem that a sequence
of  linear finite elements are designed to solve.  If we refine the graph
further, we break each spherical triangle element (in general a
D-simplex)  into a sequence of an increasing number of smaller spherical triangles,  approximated with
increasing accuracy by  flat triangles
with the linear finite element expression
(\ref{eq:ScalarFEM}).  All that remains is to numerically relax the lattice
field variables to the configuration which minimizes the action.  We
perform the relaxation using the conjugate gradient algorithm.

This procedure can be interpreted as a renormalization group flow in
the sense that we are replacing an action expressed in terms of many
linear finite elements at short distance with an action of equal value
expressed with spherical finite elements at long distance.  We are not
integrating out degrees of freedom as is done in a usual real space RG procedure,
but our theory is gaussian so there are non-linearities to take
into account. This approach  is not limited to the Riemann sphere.  In general, one
should be able to construct a ``geodesic triangular element'' on a
manifold $({\cal M},g)$ by laying down a simplicial mesh on the
geodesic element and solving for the classical solution of Laplace's
equation on the curved space using linear finite elements.

\paragraph{Conformal Mapping Approach:}
The conformal mapping approach makes use of two mathematical
identities to simplify the problem and to avoid computing the elements
directly.  The first identity uses conformal mapping to express the
action on a triangle as an integral over the unit disk in the complex plane.  The
mapping takes place in three steps.  First, stenographically project
the spherical triangle onto the flat complex plane with a triangular
image bounded by 3 circular arcs.  Second, use the Schwartz triangle
function \cite{conformal} (a generalization of the
Schwarz-Christoffel transformation for circular arc polygons) to map the
circular arc triangle onto the upper half plane.  Third, use the
M\"obius transformation to map the upper half plane onto the unit
disk.  Only the Schwartz triangle map is nontrivial, involving  
%The vertices of
%the triangle can be mapped to angles $0, 2\pi/3,4\pi/3$ on the
%unit disk.  
 ratios of hypergeometric functions that map
the  linear interpolation  of the geodesics on the spherical triangle
to non-uniform distributions on the boundaries of the disk with singularties at the vertices. Nonetheless, the interpolated boundary
data, $\tilde{\phi}(\xi(\theta))$, on the unit circle can be computed numerically
by a look up table on the original data.

The second identity uses  the divergence theorem and Green's theorem  to rewrite
the action $S_\Delta$ as an integral over the boundary of the disk.   First the divergence theorem expresses the action as
boundary value problem,
\begin{equation}
S_\Delta = \int_V d^2x (\nabla\phi)^2 = \int_{\partial V} dl
\phi\hat{n} \cdot \nabla\phi - \int_V d^2x \phi \nabla^2 \phi \; ,
\end{equation}
since the second term vanishes by the equations of motion.   Next
we can use  Green's theorem  to rewrite the solution in terms of the
boundary values,
\begin{equation}
\phi(x) = \int_{\partial V}dl' \tilde{\phi}(x') \; \hat{n'}\cdot
\nabla' G(x',x) \; .
\end{equation}
Combining these equations for the unit disk leads to a remarkable
closed form as a double boundary intergral,
\begin{equation}
S_\Delta = \frac{1}{4\pi}\int_0^{2\pi} d\theta_1 \int_0^{2\pi} d\theta_2 \frac{
  \left(
    \widetilde{\phi}(\xi(\theta_1))-\widetilde{\phi}(\xi(\theta_2))\right)^2}{1-\cos(\theta_1-\theta_2)}
\; ,
\label{eq:doubleIID}
\end{equation}
where $\xi(\theta)$ is the inverse mapping function from the boundary
of the unit disk back to the boundary of the spherical triangle.    These integrations
can be done numerically using a well constructed numerical adaptive
integrator. The crucial feature in this identity (\ref{eq:doubleIID}) is the
inclusion  of local terms, $\widetilde{\phi}^2(\xi(\theta_1))$ and $\widetilde{\phi}^2(\xi(\theta_2))$, in the integrand that renders it non-singular at  $\theta_1
= \theta_2$. Our derivation of this ``regulator'' relied on the use of 
confromality and vertex functions for the bosonic string theory.

\subsection{Spectrum of Spherical Finite Element Laplacian}
\label{subsec:Mass}
When computing spectra of operators on curved manifolds, one must
solve the generalized eigenvalue problem,
\begin{equation}
K_{ij}\phi^{(n)}_j = \lambda_n \sqrt{g_i}
\phi^{(n)}_i \; ,
\end{equation}
with eigenfunctions  that are 
orthogonal with respect to the correct measure, $\sum_i\sqrt{g_i}
\phi^{(n)}_i \phi^{(m)}_i  =\delta_{nm}$.  
The discrete measure, $\sqrt{g_i}$, may be chosen in a variety of ways.  One approach is
to take the area at a site to be the area of the Voronoi dual cell.  Another approach is to
construct the finite element expression which gives a non-local mass term:
$\int d^2x \sqrt{g(x)} \phi(x)^2 \rightarrow \mu_{ij} \phi_i \phi_j$.
We refer to this as the finite element method (FEM)  mass term.  In our
results, we consider four spectra: the flat finite element with
Voronoi area term, the flat finite element with FEM area term, the
spherical finite element with Voronoi area term, and the spherical
finite element with FEM area term.

\begin{figure}
\begin{centering}
\includegraphics[width=0.45\textwidth]{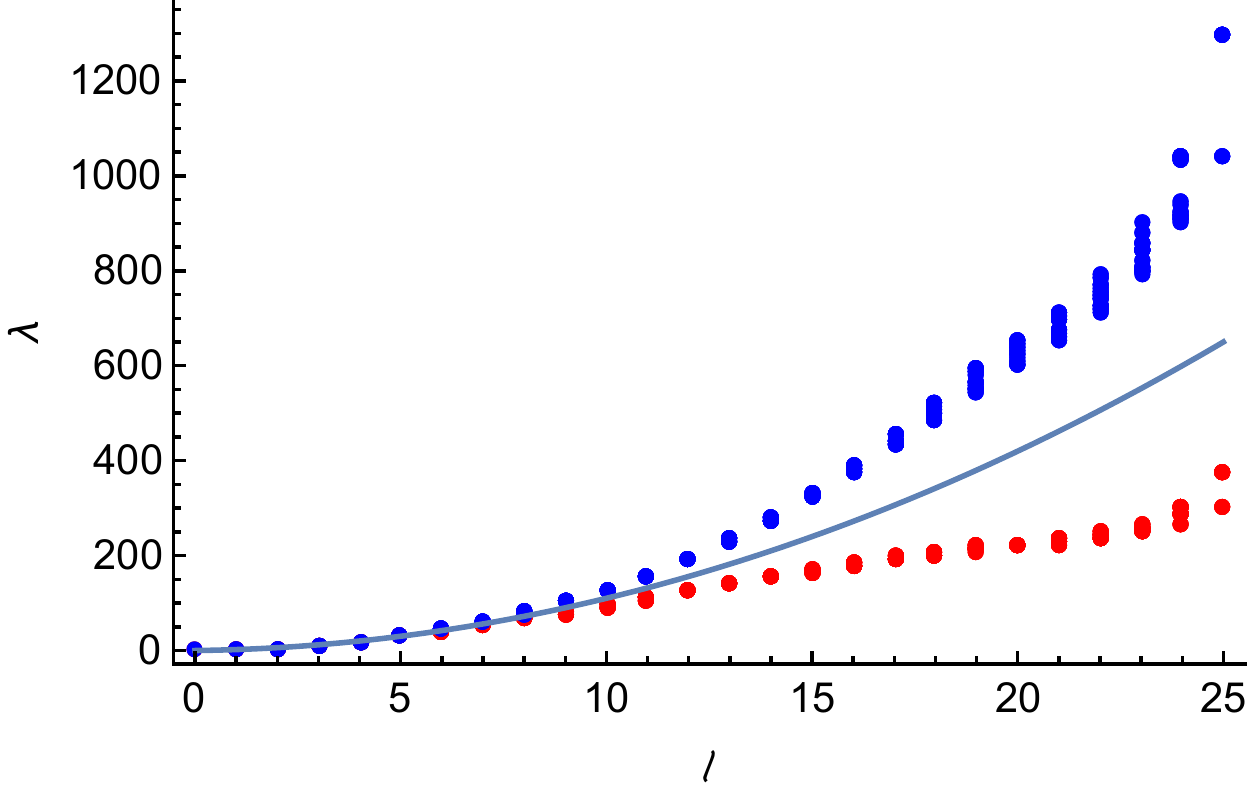}
\hskip 0.75 cm
\includegraphics[width=0.45\textwidth]{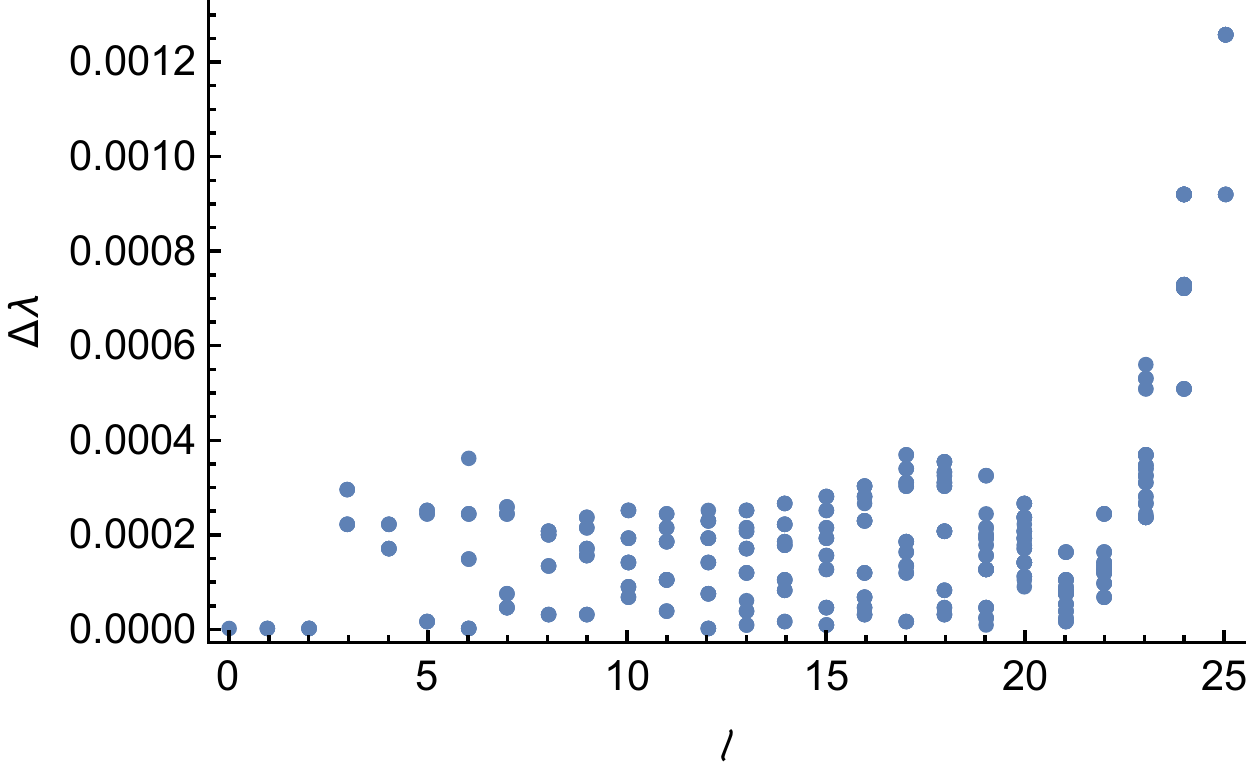}
\caption{Laplace spectra at refinement $s=8$ plotted against angular momentum quantum number, $l$.  Left: Spherical FEM spectrum using Voronoi area term (red) and FEM area term (blue).  Right: Difference in eigenvalues between flat FEM and spherical FEM; $\Delta\lambda = |\lambda_{\text{flat}} - \lambda_{\text{spherical}}|/\lambda_{\text{flat}}$.} \label{fig:spec}
\end{centering}
\end{figure}

Fig.~\ref{fig:spec} is a  plot of the complete set of
eigenvalues $2l + 1$ against the principle quantum number $l$
for a refinement of $s=8$.  On the left plot, one sees a striking
difference between the FEM area term and Voronoi area term.  The
spectrum with FEM area term converges from above, which corresponds to
a suppression of high angular momentum degrees of freedom at finite
lattice spacing, while the spectrum with Voronoi area term converges
from below, which corresponds to an enhancement of high angular
momentum degrees of freedom.  The right plot compares
the spherical element and flat element spectra, which are essentially
the same, so on this plot they merge into single circles one on top
of the  another.

\begin{figure}
\begin{centering}
\includegraphics[width=35pc]{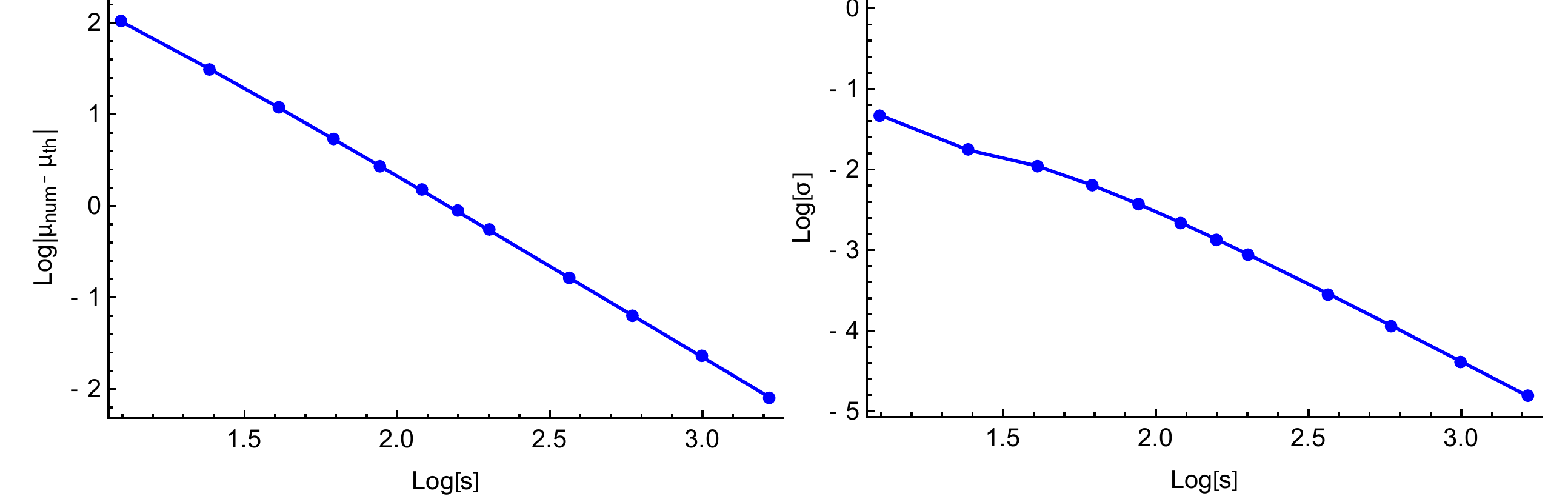}
\caption{ Convergence of spherical FEM spectrum with $s$.  We examine
  the eleven eigenvalues in the $l = 5$ level.  Left: Convergence
  of the mean of the level.  Right: Convergence of the standard deviation of level.  Both the mean and standard deviation converge like $1/s^2$.}
\label{fig:conv}
\end{centering}
\end{figure}

Fig.~\ref{fig:conv} demonstrates the convergence of the
$2 l + 1  = 11 $ eigenvalues   for $l =5$ as the lattice spacing is taken to
zero for our spherical finite element. ( Of course a very similar plot 
holds for the linear finite elements spectra.) 
In the left plot, we show the error  of the level mean from the
$2 l + 1$ eigenvalues relative to the exact  continuum value of
$l (l + 1)$.  In the right plot we show
the standard deviation of the splitting of the $m = -l, \cdots
l$  eigenvalues relative to their mean.   Both
plots are plotted against the refinement, $s$, on a log-log scale
to demonstrate convergence as $O(1/s^2)$ for sufficiently large $s$.

Finally, we note that because of the conformality of the 2D free Laplacian,
it is possible to re-interpret the spherical element as
a new element mapped onto the  original   flat simplicial complex.  On each spherical triangle, the
radial projection from the barycentric co-ordinates $\xi^1, \xi^2$ to
the spherical co-ordinates, $x = (\phi, \theta)$, maps the secants to great circles and the
metric is given by the flat metric $g_{ij}$ in Eq.~\ref{eq:SimplicialAction}, up to a
conformal factor: $g_{\mu \nu} = e^{ \sigma(x)} g_{ij}$. In
2D, the conformal factor cancels in $\sqrt{g(x)} g^{\mu
  \nu}(x)$ so we have the identity,
\be
\int_{\Delta} d^2 x \sqrt{g(x)} g^{\mu \nu}(x)
\partial_\mu \phi(x) \partial_\nu \phi(x) = \int_{\Delta} d^2 \xi \sqrt{\det[g_{ij}]} g^{ij}
\partial_i \phi(\xi) \partial_j\phi(\xi)  \; .
\ee
Consequently the  spherical elements are equivalent  to a new
element basis on the simplicial complex with flat
triangles, conforming to linearity in arclength on the projected  great circles.  In higher dimensions the conformal
 factor does not cancel,  so one cannot interpret the spherical elements in this way; nonetheless they may still be constructed and used as elements on the original manifold.

%%%%%%%%%%%%%%%%%%%%%%%%%%%%%%%%%%%%%%%%%%%%%%%%%%%%%%%%%%%%%%%%%%%%%%%%%%%%%%%%%%%%%%%%%%%%%%%%%%%%%%%%%%
\section{Conclusion}

We report on a new formulation of lattice field theory  suited to
smooth Riemann manifolds. The 2D scalar field theory on the Riemann sphere
is now under control at the Wilson-Fischer fixed point.  The 
accuracy of the Binder cumulants is comparable to previous
numeric and analytic results. The numerical precision was
dependent upon correcting the linear finte elements by a 
one loop  counter term explicitly evaluated on the simplicial complex.  The foundation for Dirac fermions on
 simplicial lattices has also been laid down.  The spectrum and
 wavefunctions for free scalars and Fermions were numerically
show to  converge to the continuum
as expected for the Riemann $\mathbb S^2$ sphere.

For the both the scalar and the Dirac simplicial lattice actions, we
note that strict adherence to linear FEM is not effective.   Instead we 
sought a new formulation which we refer to as {\bf QFE} for Quantum Finite
Elements.  While we report only on numerical tests on the Riemann
$\mathbb S^2$ sphere, the basic methods will apply more generally.
We have now developed all the necessary tools and software to  test
QFE on the radial quantization of the 3D Ising CFT at the
Wilson-Fisher fixed point.  In the future we will 
include non-Abelian gauge theories. We anticipate no fundamental
barriers  for super renormalizable field theories but asymptotically free
gauge theories in 4D will most likely require more
fundamental advances.  On the software level, we plan to extend the
SciDAC 
 data parallel software for Lattice Field Theory to simplicial lattices,  clearly a
prerequisite  to robust numerical  simulation for a large
range of  applications to  the physics  of quantum fields on curved manifolds.

\section*{Acknowledgments}

R.C.B acknowledges useful  discussions with John
Cardy, Ami Katz and Martin L\"uscher. R.C.B. and E.W. were supported by DOE grant DE-SC0010025. R.C.B. and
G.F.  thank the Aspen Center for Physics, which is supported by
National Science Foundation grant PHY-1066293 and thank the KITP, Santa Barbara, supported in part by the National Science Foundation under Grant No. NSF PHY11-25915.

\bibliographystyle{JHEP}
\bibliography{QFE}

\end{document}